\documentclass[aps,prl,reprint,notitlepage,showpacs,nofootinbib,superscriptaddress,twocolumn
,nofloats,preprintnumbers]{revtex4-1}
\usepackage{graphicx}
\usepackage[utf8]{inputenc} 
\usepackage{amsmath}
\usepackage{amssymb}
\usepackage{ulem}
\usepackage{float}
\usepackage{comment}
\usepackage{slashed}

\usepackage[T1]{fontenc} 
\usepackage{parskip}
\usepackage{bm}         
\usepackage{array}
\usepackage{multirow}   
\usepackage{booktabs}   

\usepackage{textcomp}
\usepackage{amssymb}
\usepackage{ulem}
\usepackage{cancel}

\usepackage{enumerate}
\usepackage{caption}
\usepackage{subcaption}
\usepackage{ragged2e}
 \DeclareCaptionJustification{justified}{\justifying}
\usepackage{lineno}

\usepackage{amsmath}
\usepackage{amssymb}
\usepackage{indentfirst}
\usepackage[usenames,dvipsnames]{xcolor}
\usepackage[colorlinks=true,citecolor=darkred,urlcolor=darkred, pdfborder={0 0 0}]{hyperref}
\definecolor{darkred}{rgb}{0.6,0,0}

\definecolor{linkcolor}{rgb}{0,0,0.5}
 
\newcommand {\ignore}[1]{}

\definecolor{bostonuniversityred}{rgb}{0.8, 0.0, 0.0}

\def\gsim{\raise0.3ex\hbox{$\;>$\kern-0.75em\raise-1.1ex\hbox{$\sim\;$}}}
\def\lsim{\raise0.3ex\hbox{$\;<$\kern-0.75em\raise-1.1ex\hbox{$\sim\;$}}}

\usepackage{soul}
\definecolor{mightnightblue}{RGB}{25,25,112}

\definecolor{brown}{rgb}{0.59, 0.29, 0.0}

\newcommand{\wimpy}{\texttt{WIMpy\_NREFT}}

\def\21{$\mathrm{SU(2)_L \otimes U(1)_Y}$}

\begin{document}
\preprint{APS/123-QED}

\title{Results on photon-mediated dark matter-nucleus interactions from the PICO-60 C$_{3}$F$_{8}$ bubble chamber}

\author{B.~Ali}
\affiliation{Institute of Experimental and Applied Physics, Czech Technical University in Prague, Prague, Cz-12800, Czech Republic}

\author{I.~J.~Arnquist}
\affiliation{Pacific Northwest National Laboratory, Richland, Washington 99354, USA}

\author{D.~Baxter}
\affiliation{Fermi National Accelerator Laboratory, Batavia, Illinois 60510, USA}

\author{E.~Behnke}
\affiliation{Department of Physics, Indiana University South Bend, South Bend, Indiana 46634, USA}

\author{M.~Bressler}
\affiliation{Department of Physics, Drexel University, Philadelphia, Pennsylvania 19104, USA}

\author{B.~Broerman}
\affiliation{Department of Physics, Queen's University, Kingston, K7L 3N6, Canada}

\author{C.~J.~Chen}
\affiliation{
Department of Physics and Astronomy, Northwestern University, Evanston, Illinois 60208, USA}

\author{K.~Clark}
\affiliation{Department of Physics, Queen's University, Kingston, K7L 3N6, Canada}

\author{J.~I.~Collar}
\affiliation{Enrico Fermi Institute, KICP, and Department of Physics,
University of Chicago, Chicago, Illinois 60637, USA}

\author{P.~S.~Cooper}
\affiliation{Fermi National Accelerator Laboratory, Batavia, Illinois 60510, USA}

\author{C.~Cripe}
\affiliation{Department of Physics, Indiana University South Bend, South Bend, Indiana 46634, USA}

\author{M.~Crisler}
\affiliation{Fermi National Accelerator Laboratory, Batavia, Illinois 60510, USA}

\author{C.~E.~Dahl}
\affiliation{
Department of Physics and Astronomy, Northwestern University, Evanston, Illinois 60208, USA}
\affiliation{Fermi National Accelerator Laboratory, Batavia, Illinois 60510, USA}

\author{M.~Das}
\affiliation{High Energy Nuclear \& Particle Physics Division, Saha Institute of Nuclear Physics, Kolkata, India}

\author{D.~Durnford}
\affiliation{Department of Physics, University of Alberta, Edmonton, T6G 2E1, Canada}

\author{S.~Fallows}
\affiliation{Department of Physics, University of Alberta, Edmonton, T6G 2E1, Canada}

\author{J.~Farine}
\affiliation{School of Biological, Chemical, and Forensic Sciences, Laurentian University, Sudbury, ON P3E 2C6, Canada}
\affiliation{SNOLAB, Lively, Ontario, P3Y 1N2, Canada}
\affiliation{Department of Physics, Carleton University, Ottawa, Ontario, K1S 5B6, Canada}

\author{R.~Filgas}
\affiliation{Institute of Experimental and Applied Physics, Czech Technical University in Prague, Prague, Cz-12800, Czech Republic}

\author{A. Garc\'{\i}a-Viltres}
\email[Corresponding: ]{agarciaviltres@gmail.com}
\affiliation{Instituto de F\'isica, Universidad Nacional Aut\'onoma de M\'exico, M\'exico D.\:F. 01000, M\'exico}

\author{G.~Giroux}
\affiliation{Department of Physics, Queen's University, Kingston, K7L 3N6, Canada}

\author{O.~Harris}
\affiliation{Northeastern Illinois University, Chicago, Illinois 60625, USA}

\author{T.~Hillier}
\affiliation{School of Biological, Chemical, and Forensic Sciences, Laurentian University, Sudbury, ON P3E 2C6, Canada}
\affiliation{SNOLAB, Lively, Ontario, P3Y 1N2, Canada}

\author{E.~W.~Hoppe}
\affiliation{Pacific Northwest National Laboratory, Richland, Washington 99354, USA}

\author{C.~M.~Jackson}
\affiliation{Pacific Northwest National Laboratory, Richland, Washington 99354, USA}

\author{M.~Jin}
\affiliation{
Department of Physics and Astronomy, Northwestern University, Evanston, Illinois 60208, USA}

\author{C.~B.~Krauss}
\affiliation{Department of Physics, University of Alberta, Edmonton, T6G 2E1, Canada}

\author{V.~Kumar}
\affiliation{High Energy Nuclear \& Particle Physics Division, Saha Institute of Nuclear Physics, Kolkata, India}

\author{M.~Laurin}
\affiliation{D\'epartement de Physique, Universit\'e de Montr\'eal, Montr\'eal, H3C 3J7, Canada}

\author{I.~Lawson}
\affiliation{School of Biological, Chemical, and Forensic Sciences, Laurentian University, Sudbury, ON P3E 2C6, Canada}
\affiliation{SNOLAB, Lively, Ontario, P3Y 1N2, Canada}

\author{A.~Leblanc}
\affiliation{School of Biological, Chemical, and Forensic Sciences, Laurentian University, Sudbury, ON P3E 2C6, Canada}

\author{H.~Leng}
\affiliation{Materials Research Institute, Penn State, University Park, Pennsylvania 16802, USA}

\author{I.~Levine}
\affiliation{Department of Physics, Indiana University South Bend, South Bend, Indiana 46634, USA}

\author{C.~Licciardi}
\affiliation{School of Biological, Chemical, and Forensic Sciences, Laurentian University, Sudbury, ON P3E 2C6, Canada}
\affiliation{SNOLAB, Lively, Ontario, P3Y 1N2, Canada}
\affiliation{Department of Physics, Carleton University, Ottawa, Ontario, K1S 5B6, Canada}

\author{S.~Linden}
\affiliation{SNOLAB, Lively, Ontario, P3Y 1N2, Canada}

\author{P.~Mitra}
\affiliation{Department of Physics, University of Alberta, Edmonton, T6G 2E1, Canada}

\author{V.~Monette}
\affiliation{D\'epartement de Physique, Universit\'e de Montr\'eal, Montr\'eal, H3C 3J7, Canada}

\author{C.~Moore}
\affiliation{Department of Physics, Queen's University, Kingston, K7L 3N6, Canada}

\author{R.~Neilson}
\affiliation{Department of Physics, Drexel University, Philadelphia, Pennsylvania 19104, USA}

\author{A.~J.~Noble}
\affiliation{Department of Physics, Queen's University, Kingston, K7L 3N6, Canada}

\author{H.~Nozard}
\affiliation{D\'epartement de Physique, Universit\'e de Montr\'eal, Montr\'eal, H3C 3J7, Canada}

\author{S.~Pal}
\affiliation{Department of Physics, University of Alberta, Edmonton, T6G 2E1, Canada}

\author{M.-C.~Piro}
\affiliation{Department of Physics, University of Alberta, Edmonton, T6G 2E1, Canada}

\author{A.~Plante}
\affiliation{D\'epartement de Physique, Universit\'e de Montr\'eal, Montr\'eal, H3C 3J7, Canada}

\author{S.~Priya}
\affiliation{Materials Research Institute, Penn State, University Park, Pennsylvania 16802, USA}

\author{C.~Rethmeier}
\affiliation{Department of Physics, University of Alberta, Edmonton, T6G 2E1, Canada}

\author{A.~E.~Robinson}
\affiliation{D\'epartement de Physique, Universit\'e de Montr\'eal, Montr\'eal, H3C 3J7, Canada}

\author{J.~Savoie}
\affiliation{D\'epartement de Physique, Universit\'e de Montr\'eal, Montr\'eal, H3C 3J7, Canada}

\author{A.~Sonnenschein}
\affiliation{Fermi National Accelerator Laboratory, Batavia, Illinois 60510, USA}

\author{N.~Starinski}
\affiliation{D\'epartement de Physique, Universit\'e de Montr\'eal, Montr\'eal, H3C 3J7, Canada}

\author{I.~\v{S}tekl}
\affiliation{Institute of Experimental and Applied Physics, Czech Technical University in Prague, Prague, Cz-12800, Czech Republic}

\author{D.~Tiwari}
\affiliation{D\'epartement de Physique, Universit\'e de Montr\'eal, Montr\'eal, H3C 3J7, Canada}

\author{E.~V\'azquez-J\'auregui}
\email[Corresponding: ]{ericvj@fisica.unam.mx}
\affiliation{Instituto de F\'isica, Universidad Nacional Aut\'onoma de M\'exico, M\'exico D.\:F. 01000, M\'exico}

\author{U.~Wichoski}
\affiliation{School of Biological, Chemical, and Forensic Sciences, Laurentian University, Sudbury, ON P3E 2C6, Canada}
\affiliation{SNOLAB, Lively, Ontario, P3Y 1N2, Canada}
\affiliation{Department of Physics, Carleton University, Ottawa, Ontario, K1S 5B6, Canada}

\author{V.~Zacek}
\affiliation{D\'epartement de Physique, Universit\'e de Montr\'eal, Montr\'eal, H3C 3J7, Canada}

\author{J.~Zhang}
\altaffiliation[now at ]{Argonne National Laboratory}
\affiliation{
Department of Physics and Astronomy, Northwestern University, Evanston, Illinois 60208, USA}

\collaboration{PICO Collaboration}
\noaffiliation

\date{\today}

\begin{abstract}
Many compelling models predict dark matter coupling to the electromagnetic current through higher multipole interactions, while remaining electrically neutral. Different multipole couplings have been studied, among them anapole moment, electric and magnetic dipole moments, and millicharge. This study sets limits on the couplings for these photon-mediated interactions using non-relativistic contact operators in an effective field theory framework. Using data from the PICO-60 bubble chamber leading limits for dark matter masses between 2.7 GeV/c$^2$ and 24 GeV/c$^2$ are reported for the coupling of these photon-mediated dark matter-nucleus interactions. The detector was filled with 52 kg of C$_3$F$_8$ operating at thermodynamic thresholds of 2.45 keV and 3.29 keV, reaching exposures of 1404 kg-day and 1167 kg-day, respectively.
\end{abstract}

\maketitle

\section{\label{sec:introduction}Introduction}
The identification of dark matter (DM), one of the main questions in contemporary physics, remains an elusive problem \cite{ParticleDataGroup:2020ssz,Komatsu_2009,JUNGMAN1996195,Goodman:1984dc,Cushman2013,PhysRevD.89.123521,Nobile_2016}. Direct detection experiments are low background detectors that aim to detect tiny energy deposits, O(1-100)-keV, produced by the elastic collision of Weakly Interacting Massive Particles (WIMP)~\cite{Gelmini_2017,BEREZINSKY19961,SERVANT2003391,POSPELOV2000181}. WIMPs remain promising DM candidates \cite{Roszkowski_2018,Bertone,10.1088/1361-6633/ac5754}, with several experiments setting tight constraints with cross-sections of the order of $10^{-45}$ cm$^2$ \cite{PhysRevD.96.042004,PhysRevLett.121.111302,PhysRevD.104.062005,PhysRevD.91.092004,Adhikari_2019,PhysRevD.101.062002,article,PhysRevD.103.102005,PhysRevLett.118.251301,PhysRevD.100.022001} for masses at approximately 100 GeV/c$^2$. Historically, results have been reported for couplings in terms of spin-independent (SI) and spin-dependent (SD) cross-sections \cite{PhysRevLett.128.072502,Barger2008}. As increasingly sensitive searches fail to observe convincing candidate events, interest in other interactions of DM with baryonic matter surge, well motivated by different physics scenarios. 
DM is electrically neutral, but coupling to the photon through higher multipole interactions is possible~\cite{Raby:1987ga,Raby:1987ms,BAGNASCO199499,Sigurdson2004,POSPELOV2000181,osti_1104185,Banks:2010eh,Antipin2015,PhysRevD.89.016017,PhysRevD.82.075004,BARGER201174,Nobile_2014}. Many couplings have been studied, such as anapole moment \cite{PhysRevD.101.015013,ARIAS201917,Kang_2018,PhysRevD.100.016017,Nobile_2014}, magnetic \cite{Vento2021,Hisano_2020,Sigurdson2004,Hambye2021,Nobile_2014} and electric \cite{Sigurdson2004,Hambye2021} dipole moments, and with a millicharge \cite{Bogorad2021,Li_2022,Aboubrahim2021,osti_1847202,PhysRevD.103.103523,Munnoz2018}. These photon-mediated interactions could be relevant for low WIMP masses, O(1-10)-GeV/c$^2$~\cite{Arina2021}. This work considers operators within an effective field theory as a benchmark scenario to establish limits on photon-mediated couplings using data from the PICO-60 bubble chamber. 

\section{\label{sec:pico60}PIC0-60 experiment}
The PICO-60 bubble chamber was operated two km deep underground at SNOLAB \cite{Smith:2012fq} between November 2016 and January 2017 for a first physics run and from April to June 2017 for a second run. The detector consisted of a fused silica inner vessel filled with (52.2 $\pm$ 0.5) kg of C$_3$F$_8$ in a superheated state. The inner vessel was immersed in a stainless steel pressure vessel filled with mineral oil, acting as a thermal bath and hydraulic fluid. The chamber had four cameras installed to photograph the bubble nucleation process and eight piezoelectric acoustic transducers were attached to the inner vessel to record the acoustic emissions from bubble nucleations. The first physics run had an exposure of 1167 kg-day at a 3.29-keV thermodynamic Seitz threshold, while the second had an exposure of 1404 kg-day at a 2.45-keV Seitz threshold. These two searches established leading limits on SD couplings, setting the most stringent direct-detection constraints to date on the WIMP-proton spin-dependent cross-section at $2.5 \times 10^{-41}$ cm$^2$ for a 25 GeV/c$^2$ WIMP. Details of the experimental setup, data analysis, background estimates, and WIMP search results are found in Refs.~\cite{PhysRevLett.118.251301,PhysRevD.100.022001}. The limit calculation method for SD and SI couplings, previously published~\cite{PhysRevD.100.022001} by the PICO collaboration, is followed in this work. Namely, the exclusion limits for the combined datasets are determined with a Profile Likelihood Ratio (PLR) test~\cite{PLR_2011}. Efficiency functions are obtained from calibration data using the emcee~\cite{Foreman_Mackey_2013} Markov Chain Monte Carlo (MCMC) python code package~\cite{Durnford_2022}. These functions are used to obtain the WIMP detection efficiency, for each of the couplings (expressed as functions of effective operators for the photon-mediated interactions), by integration over the nuclear recoil spectrum from an astrophysical WIMP flux for an array of potential WIMP masses. The result is a tensor containing the WIMP detection efficiency, dependent on the interaction, for each thermodynamic threshold and WIMP mass. A likelihood surface is created from this tensor at 2.45 and 3.29 keV, which is a function of the WIMP detection efficiency. This surface is convolved with a two-dimensional Gaussian function that accounts for the uncertainty of the thermodynamic thresholds. Next, the maximum of the likelihood surface for each WIMP mass is determined and used to calculate the optimal coupling. The ratio of the likelihood for a particular coupling to the maximum likelihood over all couplings is used to construct a test statistic. The exclusion curve for each of the couplings reported is obtained with toy datasets generating points in a grid of WIMP masses and couplings. A point is excluded if the evaluated PLR test statistic is larger than 90\% of toy dataset test statistics. The exclusion limits consider a local dark matter density $\rho_D=0.3 $ GeV/c$^2$/cm$^3$ within the standard halo parametrization~\cite{LEWIN199687}. The same astrophysical parameters as in the SD and SI analysis were assumed.

\section{\label{sec:photonDM} Non-relativistic effective field theory}
A non-relativistic effective field theory (NREFT) approach is suitable to extend the standard SI and SD searches. This framework allows generalizing the analysis of direct detection experiments since it considers the non-relativistic quantum mechanical operators contributing to the elastic scattering of DM with a nucleus. These interactions could provide different nuclear responses compared to the SI and SD scenarios. In this work, higher multipole interactions are studied, such as anapole moment, magnetic and electric dipole moments, and millicharge. These interactions can be generically parameterized in terms of non-relativistic effective operators~\cite{Fan_2010,fitzpatrick_effective_2013,PhysRevC.89.065501,PhysRevD.92.063515}, for which the nuclear scattering cross-sections depend on exchanged momentum, relative velocity, and nucleon and DM spins. The relevant contact operators involved in the interactions reported in this work are,

\begin{table}[H]
\begin{center}
\begin{tabular}{ l  l }
$\mathcal{O}_1=1_{\chi}1_N$ \hspace{0.7cm} & \hspace{0.4cm} $\mathcal{O}_4=\vec{S}_{\chi}\cdot\vec{S}_N$ \\
$\mathcal{O}_5=i\vec{S}_{\chi}\cdot\left(\frac{\vec{q}}{m_N}\times\vec{v}^{\perp}\right)$ \hspace{0.4cm} & \hspace{0.4cm}
$\mathcal{O}_6=\left(\vec{S}_{\chi}\cdot\frac{\vec{q}}{m_N}\right)\left(\vec{S}_N\cdot\frac{\vec{q}}{m_N}\right)$\\
$\mathcal{O}_8=\vec{S}_{\chi}\cdot\vec{v}^{\perp}$ \hspace{0.45cm} & \hspace{0.4cm} $\mathcal{O}_9=i\vec{S}_{\chi}\cdot\left(\vec{S}_N\cdot\frac{\vec{q}}{m_N}\right)$ \\
$\mathcal{O}_{11}=i\vec{S}_{\chi}\cdot\frac{\vec{q}}{m_N}$   &  \\
\end{tabular}
\label{tab:operators}
\end{center}
\end{table}
\noindent where ${m_N}$ is the nucleon mass, $\vec{q}$ is the exchanged momentum, $\vec{v}^{\perp}$ is the perpendicular component of the velocity to the momentum transfer, $\vec{S}_{\chi}$ is the spin of the DM particle, and $\vec{S}_N$ is the spin of the nucleon. The numbering scheme is followed from the NREFT definition of the operators, a result of an index in the general Lagrangian~\cite{fitzpatrick_effective_2013}.

\begin{figure*}[htpb!]
    \centering
    \includegraphics[width=\linewidth]{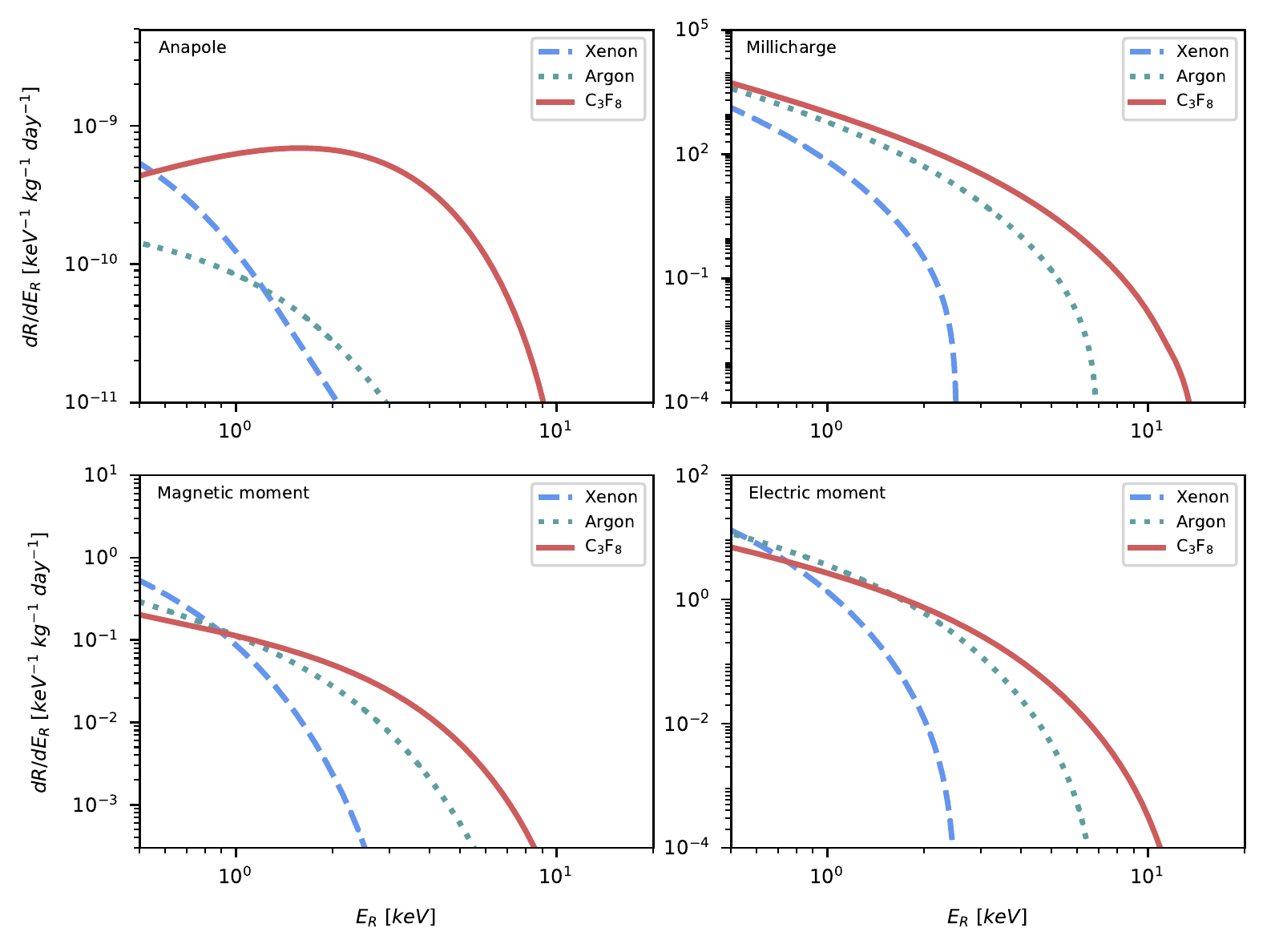}
    \caption{Scattering rates in C$_3$F$_8$ (red), xenon (dashed blue), and argon (dotted green) for a DM particle with mass of 5 GeV/c$^2$ with coupling through the anapole moment (upper left, for a coupling of $3.6\times10^{-8}$ GeV$^{-2}$), millicharge (upper right, for a coupling of $2.2\times10^{-8}\,e$), magnetic dipole moment (lower left, for a coupling of $2.8\times10^{-8}$ GeV$^{-1}$), and electric dipole moment (lower right, for a coupling of $2.8\times10^{-8}$ GeV$^{-1}$). The rates were obtained with the \wimpy\hspace{1.0mm}package~\cite{WIMpy-code}.}
    \label{fig:rates}
\end{figure*}

Photon-mediated interactions were studied using the \wimpy\hspace{1.0mm}software developed by Kavanagh et al.~\cite{WIMpy-code} which allows for the calculation of dark matter nucleus scattering rates in the framework of a non-relativistic effective field theory~\cite{Fan_2010,fitzpatrick_effective_2013}. The rate calculations for the operators involved in the interactions are in agreement with results from the dmdd (dark matter direct detection) software developed by Gluscevic et al.~\cite{Gluscevic_2015,dmdd}. The scattering rates for the operators $\mathcal{O}_1, \mathcal{O}_4, \mathcal{O}_5, \mathcal{O}_6, \mathcal{O}_8, \mathcal{O}_9,$ and $\mathcal{O}_{11}$, involved in the photon-mediated interactions, were evaluated for both software packages.
Fig.~\ref{fig:rates} shows the rates for the photon-mediated interactions, obtained with \wimpy\hspace{1.0mm}for a 5 GeV/c$^2$ DM particle. The scattering rate in fluorine for the anapole moment is significantly higher than in xenon or argon. This is primarily due to the operator $\mathcal{O}_9$ being a function of nuclear spin. In addition, the factor $\vec{q}/{m_N}$, relevant for operators $\mathcal{O}_5, \mathcal{O}_6, \mathcal{O}_9,$ and $\mathcal{O}_{11}$, results in enhanced couplings for low nuclear masses such as fluorine for WIMP masses below 20 GeV/c$^2$.

\subsection{\label{sec:anapole}Dark matter with anapole moment}
\begin{figure*}[htpb!]
    \centering
    \includegraphics[width=\linewidth]{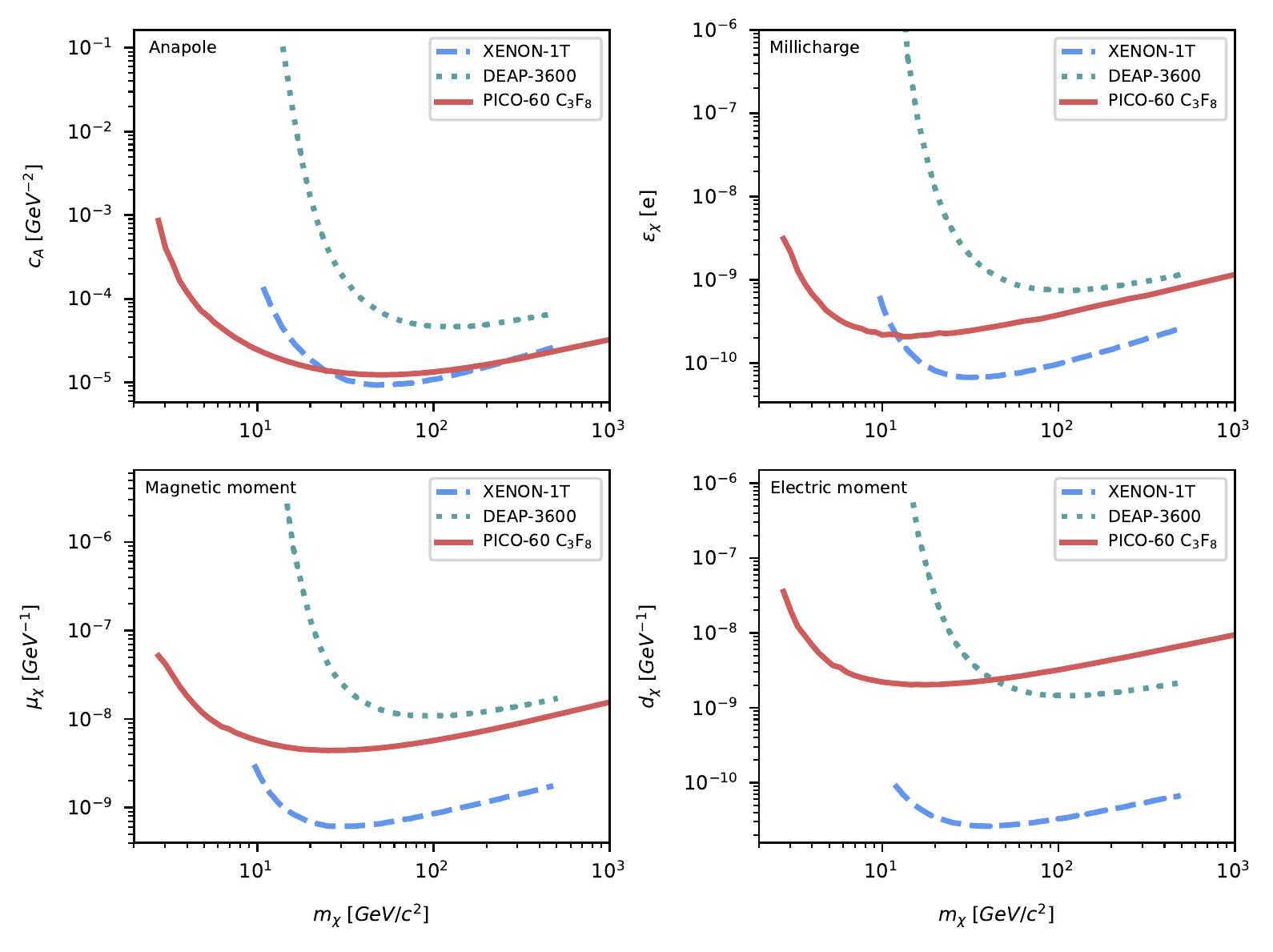}
    \caption{Exclusion limits at 90\% C.L. for the anapole moment (upper left), millicharge (upper right), magnetic dipole moment (lower left), and electric dipole moment (lower right) couplings. The limits are derived from the profile likelihood analysis of the PICO-60 C$_3$F$_8$ (red) combined blind exposure. Limits from XENON-1T (dashed blue) \cite{PhysRevD.101.063026} and DEAP-3600 (dotted green) \cite{PhysRevD.102.082001} using xenon and argon, respectively, are also shown.}
    \label{fig:all_limits}
\end{figure*}

The anapole moment is the lowest electromagnetic moment allowed for a Majorana particle. It is generated by a toroidal electric current which confines the magnetic field within a torus. It is equivalent to having a particle with a toroidal dipole moment. If the DM particle is assumed to be a Majorana fermion scattering off a nucleus via a spin-1 mediator that kinetically mixes with the photon, then the effective interaction is:
\begin{equation}
  \mathcal{L}_{\mathcal{A}}=c_{\mathcal{A}}\bar{\chi}\gamma^{\mu}\gamma^5\chi\partial^{\nu}F_{\mu\nu}\hspace{0.8mm},
\end{equation}
where the $\chi$ spinor represents the Majorana DM particle, $c_{\mathcal{A}}$ the anapole moment coupling strength and $F_{\mu\nu}$ the electromagnetic field tensor. The anapole moment has the unique feature that it interacts only with external electromagnetic currents $\mathcal{J}_{\mu}=\partial^{\nu}F_{\mu\nu}$~\cite{HO2013341}.
In the non-relativistic limit, the effective operator for anapole interactions, $\mathcal{O}_{\mathcal{A}}$, is a linear combination of the momentum-independent operator $\mathcal{O}_8$ and the momentum-dependent $\mathcal{O}_9$:
\begin{equation}
\mathcal{O}_{\mathcal{A}}=c_{\mathcal{A}}\displaystyle\sum_{N=n,p} \left( \mathcal{Q}_N\mathcal{O}_8 + g_N\mathcal{O}_9\right)\hspace{0.8mm},
\end{equation}
where $\mathcal{Q}_N$ is the nucleon charge ($\mathcal{Q}_p=e$, $\mathcal{Q}_n=0$) while $g_N$ is the nucleon g-factor ($g_p=5.59$ and $g_n=-3.83$). This interaction is expressed as $\mathcal{O}_{\mathcal{A}}=c_{\mathcal{A}}[e\mathcal{O}_8+(g_p+g_n)\mathcal{O}_9]$ for C$_3$F$_8$.  
Fig.~\ref{fig:all_limits} (upper left) shows the coupling for DM interacting through the anapole moment. The 90\% C.L. limits on the coupling from the profile likelihood analysis of the PICO-60
C$_3$F$_8$ combined blind exposure is shown and compared to results from the XENON-1T \cite{PhysRevD.101.063026} and DEAP-3600 experiments \cite{PhysRevD.102.082001}. XENON-1T and DEAP-3600 are leading experiments for SI interactions with noble liquids, using xenon and argon, respectively. PICO-60 is the leading experiment for SD interactions, using a fluorine target.

\subsection{\label{sec:anapole}Dark matter with magnetic dipole moment}
Contact interactions ($\lvert \vec{q} \rvert \ll m_{\phi}$), where $m_{\phi}$ is the mass of the mediator, are independent of the exchanged momentum; however, long-range interactions ($\lvert \vec{q} \rvert  \gg m_{\phi}$) are enhanced at small momentum transfer. Examples of long-range interactions are DM with electric or magnetic dipole moments and millicharged DM. These arise from the exchange of a massless mediator, where the propagator term enhances the interaction. Considering the DM particle as a Dirac fermion acquiring a magnetic dipole moment, the effective interaction is given by:
\begin{equation}
 \mathcal{L}_{\mathcal{M}\mathcal{D}}=\frac{\mu_{\chi}}{2}\bar{\chi}\sigma^{\mu\nu}\chi F_{\mu\nu}\hspace{0.8mm},
\end{equation}
 where the spinor $\chi$ represents the Dirac DM particle, $\mu_\chi$ is the magnetic moment coupling, and $\sigma^{\mu\nu}=\frac{i}{2}[\gamma^{\mu},\gamma^{\nu}]$. Similar to the anapole moment scenario, the non-relativistic shape of the effective operator for magnetic dipole interactions, $\mathcal{O}_{\mathcal{M}\mathcal{D}}$, can be expressed in terms of contact operators in the NREFT. $\mathcal{O}_{\mathcal{M}\mathcal{D}}$ depends on the operators $\mathcal{O}_1$, $\mathcal{O}_4$, $\mathcal{O}_5$, and $\mathcal{O}_6$ and is expressed as follows:
 \begin{equation}
 \begin{split}
 \mathcal{O}_{\mathcal{M}\mathcal{D}} =\hspace{1mm} 2e\mu_{\chi}\displaystyle\sum_{N=n,p} & \left[\mathcal{Q}_Nm_N\mathcal{O}_1+4\mathcal{Q}_N\frac{m_{\chi}m_N}{q^2}\mathcal{O}_5\right.\\ + & \left.2g_Nm_{\chi}(\mathcal{O}_4-\frac{1}{q^2}\mathcal{O}_6)\right].
 \end{split}
\end{equation}

Fig.~\ref{fig:all_limits} (lower left) presents the 90\% C.L. limits on the coupling for DM interacting through the magnetic dipole moment. 

\subsection{\label{sec:anapole}Dark matter with electric dipole moment}
Likewise, assuming a Dirac fermion as the DM particle acquiring an electric dipole moment, the effective Lagrangian for the coupling can be written as:
\begin{equation}
 \mathcal{L}_{\mathcal{E}\mathcal{D}}= \frac{d_{\chi}}{2}i\bar{\chi}\sigma^{\mu\nu}\gamma^5\chi F_{\mu\nu}\hspace{0.8mm},
\end{equation}
where $d_\chi$ is the electric dipole moment coupling. A DM particle with a permanent electric dipole moment must have a non-zero spin, and $d_{\chi}$ satisfies time-reversal and parity violation~\cite{Sigurdson2004}. The non-relativistic operator participating in this interaction, $\mathcal{O}_{\mathcal{E}\mathcal{D}}$, is a function of the $\mathcal{O}_{11}$ operator. It is expressed as:
\begin{equation}
   \mathcal{O}_{\mathcal{E}\mathcal{D}}=2ed_{\chi}\frac{1}{q^2}\mathcal{O}_{11}.
\end{equation}
Fig.~\ref{fig:all_limits} (lower right) shows the coupling for DM interacting through the electric dipole moment (90\% C.L. limits).

\subsection{\label{sec:anapole}Dark matter with millicharge}
Millicharged particles have attracted interest since they represent elegant extensions to the Standard Model \cite{HOLDOM1986196,PhysRevD.92.035014,PhysRevD.75.115001}. A millicharged DM particle would carry a fraction of the electron charge and many searches have been performed \cite{LDMX:2019,PhysRevD.101.063026,PhysRevD.85.101302,PRXQuantum.3.010330,Magill,PhysRevD.100.015043,osti_1745080,Hongwan,HAAS2015117,PhysRevLett.122.071801}. Considering a Dirac fermion, the interaction Lagrangian of the millicharged DM is given by:
\begin{equation}
\mathcal{L}_{\mathcal{M}}=e\epsilon_{\chi}A_{\mu}\bar{\chi}\gamma^{\mu}\chi,
\end{equation}
where $A_{\mu}$ is the SM photon and $\epsilon_{\chi}$ is the millicharge (a fraction of the electron charge $e$). The non-relativistic millicharge operator, $\mathcal{O}_{\mathcal{M}}$, is only a function of the $\mathcal{O}_{1}$ operator but with a $q^2$ dependence:
\begin{equation}
   \mathcal{O}_{\mathcal{M}}=e^2\epsilon_{\chi}\frac{1}{q^2}\mathcal{O}_1.
\end{equation}

Fig.~\ref{fig:all_limits} (upper right) presents the 90\% C.L. limits on the coupling for millicharged DM.

\section{\label{sec:conclusions}Conclusions}
The results presented in this work show the excellent physics reach of the bubble chamber technology using fluorine targets. World-leading limits for the coupling of photon-mediated DM interactions for masses from 2.7 GeV/c$^2$ and up to 24 GeV/c$^2$ are reported. The analysis was performed using a non-relativistic effective field theory to determine the coupling strength of the effective contact interaction operators. Assuming DM is a fermion with electromagnetic moments, the lowest order electromagnetic interaction is through the magnetic or electric dipole moments. Analysis from the PICO-60 bubble chamber sets leading limits for these couplings, as low as $2.1\times10^{-9}$ GeV$^{-1}$ for masses between 2.7 GeV/c$^2$ and 11.7 GeV/c$^2$ (electric) and $5.8\times10^{-9}$ GeV$^{-1}$ between 3 GeV/c$^2$ and 9.5 GeV/c$^2$ (magnetic). Furthermore, the only possible electromagnetic moment for a Majorana fermion is the anapole moment since the magnetic and electric dipole moments vanish. The PICO-60 experiment sets leading limits for masses between 2.7 GeV/c$^2$ and 24 GeV/c$^2$ and above 265 GeV/c$^2$ with couplings as low as $1.4\times10^{-5}$ GeV$^{-2}$. Lastly, millicharged particles are theoretically well-motivated to account for a fraction of the DM. Leading couplings as low as $2.1\times10^{-10} e$ for masses between 2.7 GeV/c$^2$ and 12 GeV/c$^2$ are obtained with data from the PICO-60 detector. The couplings reported are the strongest limits set for photon-mediated DM interactions in the low mass WIMP range (2.7-24 GeV/c$^2$). 

\section{Acknowledgements}
\begin{acknowledgments}
The PICO collaboration wishes to thank SNOLAB and its staff for support through underground space, logistical and technical services. SNOLAB operations are supported by the Canada Foundation for Innovation and the Province of Ontario Ministry of Research and Innovation, with underground access provided by Vale at the Creighton mine site. We wish to acknowledge the support of the Natural Sciences and Engineering Research Council of Canada (NSERC) and the Canada Foundation for Innovation (CFI) for funding, and the Arthur B. McDonald Canadian Astroparticle Physics Research Institute. We acknowledge that this work is supported by the National Science Foundation (NSF) (Grant 0919526, 1506337, 1242637, and 1205987), by the U.S.\ Department of Energy (DOE) Office of Science, Office of High Energy Physics (grants No. DE-SC0017815 and DE-SC-0012161), by the DOE Office of Science Graduate Student Research (SCGSR) award, by the Department of Atomic Energy (DAE), Government of India, under the Centre for AstroParticle Physics II project (CAPP-II) at the Saha Institute of Nuclear Physics (SINP), and Institutional support of IEAP CTU (DKRVO). This work is also supported by the German-Mexican research collaboration grant SP 778/4-1 (DFG) and 278017 (CONACYT), the project CONACYT CB-2017-2018/A1-S-8960,  DGAPA UNAM grant PAPIIT-IN108020, and Fundación Marcos Moshinsky. This work is partially supported by the Kavli Institute for Cosmological Physics at the University of Chicago through NSF grants 1125897 and 1806722, and an endowment from the Kavli Foundation and its founder Fred Kavli. We also wish to acknowledge the support from Fermi National Accelerator Laboratory under Contract No.\:DE-AC02-07CH11359, and from Pacific Northwest National Laboratory, which is operated by Battelle for the U.S.\ Department of Energy under Contract No.\:DE-AC05-76RL01830. We also thank Compute Canada (\url{www.computecanada.ca}) and the Centre for Advanced Computing, ACENET, Calcul Qu\'ebec, Compute Ontario, and WestGrid for computational support. The work of M.~Bressler is supported by the Department of Energy Office of Science Graduate Instrumentation Research Award (GIRA). The work of D.~Durnford is supported by the NSERC Canada Graduate Scholarships - Doctoral program (CGSD). IUSB wishes to acknowledge the work of D. Marizata.

\end{acknowledgments}
\bibliographystyle{apsrev4-1}
\bibliography{apssamp}

\begin{thebibliography}{78}%
\makeatletter
\providecommand \@ifxundefined [1]{%
 \@ifx{#1\undefined}
}%
\providecommand \@ifnum [1]{%
 \ifnum #1\expandafter \@firstoftwo
 \else \expandafter \@secondoftwo
 \fi
}%
\providecommand \@ifx [1]{%
 \ifx #1\expandafter \@firstoftwo
 \else \expandafter \@secondoftwo
 \fi
}%
\providecommand \natexlab [1]{#1}%
\providecommand \enquote  [1]{``#1''}%
\providecommand \bibnamefont  [1]{#1}%
\providecommand \bibfnamefont [1]{#1}%
\providecommand \citenamefont [1]{#1}%
\providecommand \href@noop [0]{\@secondoftwo}%
\providecommand \href [0]{\begingroup \@sanitize@url \@href}%
\providecommand \@href[1]{\@@startlink{#1}\@@href}%
\providecommand \@@href[1]{\endgroup#1\@@endlink}%
\providecommand \@sanitize@url [0]{\catcode `\\12\catcode `\$12\catcode
  `\&12\catcode `\#12\catcode `\^12\catcode `\_12\catcode `\%12\relax}%
\providecommand \@@startlink[1]{}%
\providecommand \@@endlink[0]{}%
\providecommand \url  [0]{\begingroup\@sanitize@url \@url }%
\providecommand \@url [1]{\endgroup\@href {#1}{\urlprefix }}%
\providecommand \urlprefix  [0]{URL }%
\providecommand \Eprint [0]{\href }%
\providecommand \doibase [0]{http://dx.doi.org/}%
\providecommand \selectlanguage [0]{\@gobble}%
\providecommand \bibinfo  [0]{\@secondoftwo}%
\providecommand \bibfield  [0]{\@secondoftwo}%
\providecommand \translation [1]{[#1]}%
\providecommand \BibitemOpen [0]{}%
\providecommand \bibitemStop [0]{}%
\providecommand \bibitemNoStop [0]{.\EOS\space}%
\providecommand \EOS [0]{\spacefactor3000\relax}%
\providecommand \BibitemShut  [1]{\csname bibitem#1\endcsname}%
\let\auto@bib@innerbib\@empty
\bibitem [{\citenamefont {Zyla}\ \emph {et~al.}(2020)\citenamefont {Zyla} \emph
  {et~al.}}]{ParticleDataGroup:2020ssz}%
  \BibitemOpen
  \bibfield  {author} {\bibinfo {author} {\bibfnamefont {P.~A.}\ \bibnamefont
  {Zyla}} \emph {et~al.} (\bibinfo {collaboration} {Particle Data Group}),\
  }\href {\doibase 10.1093/ptep/ptaa104} {\bibfield  {journal} {\bibinfo
  {journal} {PTEP}\ }\textbf {\bibinfo {volume} {2020}},\ \bibinfo {pages}
  {083C01} (\bibinfo {year} {2020})}\BibitemShut {NoStop}%
\bibitem [{\citenamefont {Komatsu}\ \emph {et~al.}(2009)\citenamefont
  {Komatsu}, \citenamefont {Dunkley}, \citenamefont {Nolta}, \citenamefont
  {Bennett}, \citenamefont {Gold}, \citenamefont {Hinshaw}, \citenamefont
  {Jarosik}, \citenamefont {Larson}, \citenamefont {Limon}, \citenamefont
  {Page}, \citenamefont {Spergel}, \citenamefont {Halpern}, \citenamefont
  {Hill}, \citenamefont {Kogut}, \citenamefont {Meyer}, \citenamefont {Tucker},
  \citenamefont {Weiland}, \citenamefont {Wollack},\ and\ \citenamefont
  {Wright}}]{Komatsu_2009}%
  \BibitemOpen
  \bibfield  {author} {\bibinfo {author} {\bibfnamefont {E.}~\bibnamefont
  {Komatsu}}, \bibinfo {author} {\bibfnamefont {J.}~\bibnamefont {Dunkley}},
  \bibinfo {author} {\bibfnamefont {M.~R.}\ \bibnamefont {Nolta}}, \bibinfo
  {author} {\bibfnamefont {C.~L.}\ \bibnamefont {Bennett}}, \bibinfo {author}
  {\bibfnamefont {B.}~\bibnamefont {Gold}}, \bibinfo {author} {\bibfnamefont
  {G.}~\bibnamefont {Hinshaw}}, \bibinfo {author} {\bibfnamefont
  {N.}~\bibnamefont {Jarosik}}, \bibinfo {author} {\bibfnamefont
  {D.}~\bibnamefont {Larson}}, \bibinfo {author} {\bibfnamefont
  {M.}~\bibnamefont {Limon}}, \bibinfo {author} {\bibfnamefont
  {L.}~\bibnamefont {Page}}, \bibinfo {author} {\bibfnamefont {D.~N.}\
  \bibnamefont {Spergel}}, \bibinfo {author} {\bibfnamefont {M.}~\bibnamefont
  {Halpern}}, \bibinfo {author} {\bibfnamefont {R.~S.}\ \bibnamefont {Hill}},
  \bibinfo {author} {\bibfnamefont {A.}~\bibnamefont {Kogut}}, \bibinfo
  {author} {\bibfnamefont {S.~S.}\ \bibnamefont {Meyer}}, \bibinfo {author}
  {\bibfnamefont {G.~S.}\ \bibnamefont {Tucker}}, \bibinfo {author}
  {\bibfnamefont {J.~L.}\ \bibnamefont {Weiland}}, \bibinfo {author}
  {\bibfnamefont {E.}~\bibnamefont {Wollack}}, \ and\ \bibinfo {author}
  {\bibfnamefont {E.~L.}\ \bibnamefont {Wright}},\ }\href {\doibase
  10.1088/0067-0049/180/2/330} {\bibfield  {journal} {\bibinfo  {journal} {The
  Astrophysical Journal Supplement Series}\ }\textbf {\bibinfo {volume}
  {180}},\ \bibinfo {pages} {330} (\bibinfo {year} {2009})}\BibitemShut
  {NoStop}%
\bibitem [{\citenamefont {Jungman}\ \emph {et~al.}(1996)\citenamefont
  {Jungman}, \citenamefont {Kamionkowski},\ and\ \citenamefont
  {Griest}}]{JUNGMAN1996195}%
  \BibitemOpen
  \bibfield  {author} {\bibinfo {author} {\bibfnamefont {G.}~\bibnamefont
  {Jungman}}, \bibinfo {author} {\bibfnamefont {M.}~\bibnamefont
  {Kamionkowski}}, \ and\ \bibinfo {author} {\bibfnamefont {K.}~\bibnamefont
  {Griest}},\ }\href {\doibase https://doi.org/10.1016/0370-1573(95)00058-5}
  {\bibfield  {journal} {\bibinfo  {journal} {Physics Reports}\ }\textbf
  {\bibinfo {volume} {267}},\ \bibinfo {pages} {195} (\bibinfo {year}
  {1996})}\BibitemShut {NoStop}%
\bibitem [{\citenamefont {Goodman}\ and\ \citenamefont
  {Witten}(1985)}]{Goodman:1984dc}%
  \BibitemOpen
  \bibfield  {author} {\bibinfo {author} {\bibfnamefont {M.~W.}\ \bibnamefont
  {Goodman}}\ and\ \bibinfo {author} {\bibfnamefont {E.}~\bibnamefont
  {Witten}},\ }\href {\doibase 10.1103/PhysRevD.31.3059} {\bibfield  {journal}
  {\bibinfo  {journal} {Phys. Rev. D}\ }\textbf {\bibinfo {volume} {31}},\
  \bibinfo {pages} {3059} (\bibinfo {year} {1985})}\BibitemShut {NoStop}%
\bibitem [{\citenamefont {Cushman}\ \emph {et~al.}(2013)\citenamefont {Cushman}
  \emph {et~al.}}]{Cushman2013}%
  \BibitemOpen
  \bibfield  {author} {\bibinfo {author} {\bibfnamefont {P.}~\bibnamefont
  {Cushman}} \emph {et~al.},\ }\href
  {https://www.researchgate.net/publication/258201295_Snowmass_CF1_Summary_WIMP_Dark_Matter_Direct_Detection}
  {\  (\bibinfo {year} {2013})}\BibitemShut {NoStop}%
\bibitem [{\citenamefont {Gresham}\ and\ \citenamefont
  {Zurek}(2014{\natexlab{a}})}]{PhysRevD.89.123521}%
  \BibitemOpen
  \bibfield  {author} {\bibinfo {author} {\bibfnamefont {M.~I.}\ \bibnamefont
  {Gresham}}\ and\ \bibinfo {author} {\bibfnamefont {K.~M.}\ \bibnamefont
  {Zurek}},\ }\href {\doibase 10.1103/PhysRevD.89.123521} {\bibfield  {journal}
  {\bibinfo  {journal} {Phys. Rev. D}\ }\textbf {\bibinfo {volume} {89}},\
  \bibinfo {pages} {123521} (\bibinfo {year} {2014}{\natexlab{a}})}\BibitemShut
  {NoStop}%
\bibitem [{\citenamefont {Nobile}\ \emph {et~al.}(2016)\citenamefont {Nobile},
  \citenamefont {Gelmini},\ and\ \citenamefont {Witte}}]{Nobile_2016}%
  \BibitemOpen
  \bibfield  {author} {\bibinfo {author} {\bibfnamefont {E.~D.}\ \bibnamefont
  {Nobile}}, \bibinfo {author} {\bibfnamefont {G.~B.}\ \bibnamefont {Gelmini}},
  \ and\ \bibinfo {author} {\bibfnamefont {S.~J.}\ \bibnamefont {Witte}},\
  }\href {\doibase 10.1088/1475-7516/2016/02/009} {\bibfield  {journal}
  {\bibinfo  {journal} {Journal of Cosmology and Astroparticle Physics}\
  }\textbf {\bibinfo {volume} {2016}},\ \bibinfo {pages} {009} (\bibinfo {year}
  {2016})}\BibitemShut {NoStop}%
\bibitem [{\citenamefont {Gelmini}(2017)}]{Gelmini_2017}%
  \BibitemOpen
  \bibfield  {author} {\bibinfo {author} {\bibfnamefont {G.~B.}\ \bibnamefont
  {Gelmini}},\ }\href {\doibase 10.1088/1361-6633/aa6e5c} {\bibfield  {journal}
  {\bibinfo  {journal} {Reports on Progress in Physics}\ }\textbf {\bibinfo
  {volume} {80}},\ \bibinfo {pages} {082201} (\bibinfo {year}
  {2017})}\BibitemShut {NoStop}%
\bibitem [{\citenamefont {Berezinsky}\ \emph {et~al.}(1996)\citenamefont
  {Berezinsky}, \citenamefont {Bottino}, \citenamefont {Ellis}, \citenamefont
  {Fornengo}, \citenamefont {Mignola},\ and\ \citenamefont
  {Scopel}}]{BEREZINSKY19961}%
  \BibitemOpen
  \bibfield  {author} {\bibinfo {author} {\bibfnamefont {V.}~\bibnamefont
  {Berezinsky}}, \bibinfo {author} {\bibfnamefont {A.}~\bibnamefont {Bottino}},
  \bibinfo {author} {\bibfnamefont {J.}~\bibnamefont {Ellis}}, \bibinfo
  {author} {\bibfnamefont {N.}~\bibnamefont {Fornengo}}, \bibinfo {author}
  {\bibfnamefont {G.}~\bibnamefont {Mignola}}, \ and\ \bibinfo {author}
  {\bibfnamefont {S.}~\bibnamefont {Scopel}},\ }\href {\doibase
  https://doi.org/10.1016/0927-6505(95)00048-8} {\bibfield  {journal} {\bibinfo
   {journal} {Astroparticle Physics}\ }\textbf {\bibinfo {volume} {5}},\
  \bibinfo {pages} {1} (\bibinfo {year} {1996})}\BibitemShut {NoStop}%
\bibitem [{\citenamefont {Servant}\ and\ \citenamefont
  {Tait}(2003)}]{SERVANT2003391}%
  \BibitemOpen
  \bibfield  {author} {\bibinfo {author} {\bibfnamefont {G.}~\bibnamefont
  {Servant}}\ and\ \bibinfo {author} {\bibfnamefont {T.~M.}\ \bibnamefont
  {Tait}},\ }\href {\doibase https://doi.org/10.1016/S0550-3213(02)01012-X}
  {\bibfield  {journal} {\bibinfo  {journal} {Nuclear Physics B}\ }\textbf
  {\bibinfo {volume} {650}},\ \bibinfo {pages} {391} (\bibinfo {year}
  {2003})}\BibitemShut {NoStop}%
\bibitem [{\citenamefont {Pospelov}\ and\ \citenamefont {{ter
  Veldhuis}}(2000)}]{POSPELOV2000181}%
  \BibitemOpen
  \bibfield  {author} {\bibinfo {author} {\bibfnamefont {M.}~\bibnamefont
  {Pospelov}}\ and\ \bibinfo {author} {\bibfnamefont {T.}~\bibnamefont {{ter
  Veldhuis}}},\ }\href {\doibase https://doi.org/10.1016/S0370-2693(00)00358-0}
  {\bibfield  {journal} {\bibinfo  {journal} {Physics Letters B}\ }\textbf
  {\bibinfo {volume} {480}},\ \bibinfo {pages} {181} (\bibinfo {year}
  {2000})}\BibitemShut {NoStop}%
\bibitem [{\citenamefont {Roszkowski}\ \emph {et~al.}(2018)\citenamefont
  {Roszkowski}, \citenamefont {Sessolo},\ and\ \citenamefont
  {Trojanowski}}]{Roszkowski_2018}%
  \BibitemOpen
  \bibfield  {author} {\bibinfo {author} {\bibfnamefont {L.}~\bibnamefont
  {Roszkowski}}, \bibinfo {author} {\bibfnamefont {E.~M.}\ \bibnamefont
  {Sessolo}}, \ and\ \bibinfo {author} {\bibfnamefont {S.}~\bibnamefont
  {Trojanowski}},\ }\href {\doibase 10.1088/1361-6633/aab913} {\bibfield
  {journal} {\bibinfo  {journal} {Reports on Progress in Physics}\ }\textbf
  {\bibinfo {volume} {81}},\ \bibinfo {pages} {066201} (\bibinfo {year}
  {2018})}\BibitemShut {NoStop}%
\bibitem [{\citenamefont {Bertone}\ \emph {et~al.}(2017)\citenamefont
  {Bertone}, \citenamefont {Bozorgnia}, \citenamefont {Kim}, \citenamefont
  {Liem}, \citenamefont {McCabe}, \citenamefont {Otten},\ and\ \citenamefont
  {Austri}}]{Bertone}%
  \BibitemOpen
  \bibfield  {author} {\bibinfo {author} {\bibfnamefont {G.}~\bibnamefont
  {Bertone}}, \bibinfo {author} {\bibfnamefont {N.}~\bibnamefont {Bozorgnia}},
  \bibinfo {author} {\bibfnamefont {J.}~\bibnamefont {Kim}}, \bibinfo {author}
  {\bibfnamefont {S.}~\bibnamefont {Liem}}, \bibinfo {author} {\bibfnamefont
  {C.}~\bibnamefont {McCabe}}, \bibinfo {author} {\bibfnamefont
  {S.}~\bibnamefont {Otten}}, \ and\ \bibinfo {author} {\bibfnamefont
  {R.}~\bibnamefont {Austri}},\ }\href {\doibase 10.1088/1475-7516/2018/03/026}
  {\bibfield  {journal} {\bibinfo  {journal} {Journal of Cosmology and
  Astroparticle Physics}\ }\textbf {\bibinfo {volume} {2018}} (\bibinfo {year}
  {2017}),\ 10.1088/1475-7516/2018/03/026}\BibitemShut {NoStop}%
\bibitem [{\citenamefont {Billard}\ \emph {et~al.}(2022)\citenamefont
  {Billard}, \citenamefont {Boulay}, \citenamefont {Cebrian}, \citenamefont
  {Covi}, \citenamefont {Fiorillo}, \citenamefont {Green}, \citenamefont
  {Kopp}, \citenamefont {Majorovits}, \citenamefont {Palladino}, \citenamefont
  {Petricca}, \citenamefont {Roszkowski},\ and\ \citenamefont
  {Schumann}}]{10.1088/1361-6633/ac5754}%
  \BibitemOpen
  \bibfield  {author} {\bibinfo {author} {\bibfnamefont {J.}~\bibnamefont
  {Billard}}, \bibinfo {author} {\bibfnamefont {M.}~\bibnamefont {Boulay}},
  \bibinfo {author} {\bibfnamefont {S.}~\bibnamefont {Cebrian}}, \bibinfo
  {author} {\bibfnamefont {L.}~\bibnamefont {Covi}}, \bibinfo {author}
  {\bibfnamefont {G.}~\bibnamefont {Fiorillo}}, \bibinfo {author}
  {\bibfnamefont {A.~M.}\ \bibnamefont {Green}}, \bibinfo {author}
  {\bibfnamefont {J.}~\bibnamefont {Kopp}}, \bibinfo {author} {\bibfnamefont
  {B.}~\bibnamefont {Majorovits}}, \bibinfo {author} {\bibfnamefont
  {K.}~\bibnamefont {Palladino}}, \bibinfo {author} {\bibfnamefont
  {F.}~\bibnamefont {Petricca}}, \bibinfo {author} {\bibfnamefont
  {L.}~\bibnamefont {Roszkowski}}, \ and\ \bibinfo {author} {\bibfnamefont
  {M.}~\bibnamefont {Schumann}},\ }\href
  {http://iopscience.iop.org/article/10.1088/1361-6633/ac5754} {\bibfield
  {journal} {\bibinfo  {journal} {Reports on Progress in Physics}\ } (\bibinfo
  {year} {2022})}\BibitemShut {NoStop}%
\bibitem [{\citenamefont {Aprile}\ \emph {et~al.}(2017)\citenamefont {Aprile}
  \emph {et~al.}}]{PhysRevD.96.042004}%
  \BibitemOpen
  \bibfield  {author} {\bibinfo {author} {\bibfnamefont {E.}~\bibnamefont
  {Aprile}} \emph {et~al.} (\bibinfo {collaboration} {XENON Collaboration}),\
  }\href {\doibase 10.1103/PhysRevD.96.042004} {\bibfield  {journal} {\bibinfo
  {journal} {Phys. Rev. D}\ }\textbf {\bibinfo {volume} {96}},\ \bibinfo
  {pages} {042004} (\bibinfo {year} {2017})}\BibitemShut {NoStop}%
\bibitem [{\citenamefont {Aprile}\ \emph {et~al.}(2018)\citenamefont {Aprile}
  \emph {et~al.}}]{PhysRevLett.121.111302}%
  \BibitemOpen
  \bibfield  {author} {\bibinfo {author} {\bibfnamefont {E.}~\bibnamefont
  {Aprile}} \emph {et~al.} (\bibinfo {collaboration} {XENON Collaboration 7}),\
  }\href {\doibase 10.1103/PhysRevLett.121.111302} {\bibfield  {journal}
  {\bibinfo  {journal} {Phys. Rev. Lett.}\ }\textbf {\bibinfo {volume} {121}},\
  \bibinfo {pages} {111302} (\bibinfo {year} {2018})}\BibitemShut {NoStop}%
\bibitem [{\citenamefont {Akerib}\ \emph {et~al.}(2021)\citenamefont {Akerib}
  \emph {et~al.}}]{PhysRevD.104.062005}%
  \BibitemOpen
  \bibfield  {author} {\bibinfo {author} {\bibfnamefont {D.~S.}\ \bibnamefont
  {Akerib}} \emph {et~al.} (\bibinfo {collaboration} {LUX Collaboration}),\
  }\href {\doibase 10.1103/PhysRevD.104.062005} {\bibfield  {journal} {\bibinfo
   {journal} {Phys. Rev. D}\ }\textbf {\bibinfo {volume} {104}},\ \bibinfo
  {pages} {062005} (\bibinfo {year} {2021})}\BibitemShut {NoStop}%
\bibitem [{\citenamefont {Schneck}\ \emph {et~al.}(2015)\citenamefont {Schneck}
  \emph {et~al.}}]{PhysRevD.91.092004}%
  \BibitemOpen
  \bibfield  {author} {\bibinfo {author} {\bibfnamefont {K.}~\bibnamefont
  {Schneck}} \emph {et~al.} (\bibinfo {collaboration} {SuperCDMS
  Collaboration}),\ }\href {\doibase 10.1103/PhysRevD.91.092004} {\bibfield
  {journal} {\bibinfo  {journal} {Phys. Rev. D}\ }\textbf {\bibinfo {volume}
  {91}},\ \bibinfo {pages} {092004} (\bibinfo {year} {2015})}\BibitemShut
  {NoStop}%
\bibitem [{\citenamefont {Adhikari}\ \emph {et~al.}(2019)\citenamefont
  {Adhikari} \emph {et~al.}}]{Adhikari_2019}%
  \BibitemOpen
  \bibfield  {author} {\bibinfo {author} {\bibfnamefont {G.}~\bibnamefont
  {Adhikari}} \emph {et~al.},\ }\href {\doibase 10.1088/1475-7516/2019/06/048}
  {\bibfield  {journal} {\bibinfo  {journal} {Journal of Cosmology and
  Astroparticle Physics}\ }\textbf {\bibinfo {volume} {2019}},\ \bibinfo
  {pages} {048} (\bibinfo {year} {2019})}\BibitemShut {NoStop}%
\bibitem [{\citenamefont {Agnes}\ \emph {et~al.}(2020)\citenamefont {Agnes}
  \emph {et~al.}}]{PhysRevD.101.062002}%
  \BibitemOpen
  \bibfield  {author} {\bibinfo {author} {\bibfnamefont {P.}~\bibnamefont
  {Agnes}} \emph {et~al.} (\bibinfo {collaboration} {DarkSide-50
  Collaboration}),\ }\href {\doibase 10.1103/PhysRevD.101.062002} {\bibfield
  {journal} {\bibinfo  {journal} {Phys. Rev. D}\ }\textbf {\bibinfo {volume}
  {101}},\ \bibinfo {pages} {062002} (\bibinfo {year} {2020})}\BibitemShut
  {NoStop}%
\bibitem [{\citenamefont {Bernabei}\ \emph {et~al.}(2018)\citenamefont
  {Bernabei} \emph {et~al.}}]{article}%
  \BibitemOpen
  \bibfield  {author} {\bibinfo {author} {\bibfnamefont {R.}~\bibnamefont
  {Bernabei}} \emph {et~al.},\ }\href {\doibase 10.15407/jnpae2018.04.307}
  {\bibfield  {journal} {\bibinfo  {journal} {Nuclear Physics and Atomic
  Energy}\ }\textbf {\bibinfo {volume} {19}},\ \bibinfo {pages} {307} (\bibinfo
  {year} {2018})}\BibitemShut {NoStop}%
\bibitem [{\citenamefont {Amar\'e}\ \emph {et~al.}(2021)\citenamefont {Amar\'e}
  \emph {et~al.}}]{PhysRevD.103.102005}%
  \BibitemOpen
  \bibfield  {author} {\bibinfo {author} {\bibfnamefont {J.}~\bibnamefont
  {Amar\'e}} \emph {et~al.},\ }\href {\doibase 10.1103/PhysRevD.103.102005}
  {\bibfield  {journal} {\bibinfo  {journal} {Phys. Rev. D}\ }\textbf {\bibinfo
  {volume} {103}},\ \bibinfo {pages} {102005} (\bibinfo {year}
  {2021})}\BibitemShut {NoStop}%
\bibitem [{\citenamefont {Amole}\ \emph {et~al.}(2017)\citenamefont {Amole}
  \emph {et~al.}}]{PhysRevLett.118.251301}%
  \BibitemOpen
  \bibfield  {author} {\bibinfo {author} {\bibfnamefont {C.}~\bibnamefont
  {Amole}} \emph {et~al.} (\bibinfo {collaboration} {PICO Collaboration}),\
  }\href {\doibase 10.1103/PhysRevLett.118.251301} {\bibfield  {journal}
  {\bibinfo  {journal} {Phys. Rev. Lett.}\ }\textbf {\bibinfo {volume} {118}},\
  \bibinfo {pages} {251301} (\bibinfo {year} {2017})}\BibitemShut {NoStop}%
\bibitem [{\citenamefont {Amole}\ \emph {et~al.}(2019)\citenamefont {Amole}
  \emph {et~al.}}]{PhysRevD.100.022001}%
  \BibitemOpen
  \bibfield  {author} {\bibinfo {author} {\bibfnamefont {C.}~\bibnamefont
  {Amole}} \emph {et~al.} (\bibinfo {collaboration} {PICO Collaboration}),\
  }\href {\doibase 10.1103/PhysRevD.100.022001} {\bibfield  {journal} {\bibinfo
   {journal} {Phys. Rev. D}\ }\textbf {\bibinfo {volume} {100}},\ \bibinfo
  {pages} {022001} (\bibinfo {year} {2019})}\BibitemShut {NoStop}%
\bibitem [{\citenamefont {Hu}\ \emph {et~al.}(2022)\citenamefont {Hu},
  \citenamefont {Padua-Arg\"uelles}, \citenamefont {Leutheusser}, \citenamefont
  {Miyagi}, \citenamefont {Stroberg},\ and\ \citenamefont
  {Holt}}]{PhysRevLett.128.072502}%
  \BibitemOpen
  \bibfield  {author} {\bibinfo {author} {\bibfnamefont {B.~S.}\ \bibnamefont
  {Hu}}, \bibinfo {author} {\bibfnamefont {J.}~\bibnamefont
  {Padua-Arg\"uelles}}, \bibinfo {author} {\bibfnamefont {S.}~\bibnamefont
  {Leutheusser}}, \bibinfo {author} {\bibfnamefont {T.}~\bibnamefont {Miyagi}},
  \bibinfo {author} {\bibfnamefont {S.~R.}\ \bibnamefont {Stroberg}}, \ and\
  \bibinfo {author} {\bibfnamefont {J.~D.}\ \bibnamefont {Holt}},\ }\href
  {\doibase 10.1103/PhysRevLett.128.072502} {\bibfield  {journal} {\bibinfo
  {journal} {Phys. Rev. Lett.}\ }\textbf {\bibinfo {volume} {128}},\ \bibinfo
  {pages} {072502} (\bibinfo {year} {2022})}\BibitemShut {NoStop}%
\bibitem [{\citenamefont {Barger}\ \emph {et~al.}(2008)\citenamefont {Barger},
  \citenamefont {Keung},\ and\ \citenamefont {Shaughnessy}}]{Barger2008}%
  \BibitemOpen
  \bibfield  {author} {\bibinfo {author} {\bibfnamefont {V.}~\bibnamefont
  {Barger}}, \bibinfo {author} {\bibfnamefont {W.-Y.}\ \bibnamefont {Keung}}, \
  and\ \bibinfo {author} {\bibfnamefont {G.}~\bibnamefont {Shaughnessy}},\
  }\href {\doibase 10.1103/PhysRevD.78.056007} {\bibfield  {journal} {\bibinfo
  {journal} {Physical Review D}\ }\textbf {\bibinfo {volume} {78}} (\bibinfo
  {year} {2008}),\ 10.1103/PhysRevD.78.056007}\BibitemShut {NoStop}%
\bibitem [{\citenamefont {Raby}\ and\ \citenamefont
  {West}(1987)}]{Raby:1987ga}%
  \BibitemOpen
  \bibfield  {author} {\bibinfo {author} {\bibfnamefont {S.}~\bibnamefont
  {Raby}}\ and\ \bibinfo {author} {\bibfnamefont {G.}~\bibnamefont {West}},\
  }\href {\doibase 10.1016/0370-2693(87)90234-6} {\bibfield  {journal}
  {\bibinfo  {journal} {Phys. Lett. B}\ }\textbf {\bibinfo {volume} {194}},\
  \bibinfo {pages} {557} (\bibinfo {year} {1987})}\BibitemShut {NoStop}%
\bibitem [{\citenamefont {Raby}\ and\ \citenamefont
  {West}(1988)}]{Raby:1987ms}%
  \BibitemOpen
  \bibfield  {author} {\bibinfo {author} {\bibfnamefont {S.}~\bibnamefont
  {Raby}}\ and\ \bibinfo {author} {\bibfnamefont {G.}~\bibnamefont {West}},\
  }\href {\doibase 10.1016/0370-2693(88)90169-4} {\bibfield  {journal}
  {\bibinfo  {journal} {Phys. Lett. B}\ }\textbf {\bibinfo {volume} {200}},\
  \bibinfo {pages} {547} (\bibinfo {year} {1988})}\BibitemShut {NoStop}%
\bibitem [{\citenamefont {Bagnasco}\ \emph {et~al.}(1994)\citenamefont
  {Bagnasco}, \citenamefont {Dine},\ and\ \citenamefont
  {Thomas}}]{BAGNASCO199499}%
  \BibitemOpen
  \bibfield  {author} {\bibinfo {author} {\bibfnamefont {J.}~\bibnamefont
  {Bagnasco}}, \bibinfo {author} {\bibfnamefont {M.}~\bibnamefont {Dine}}, \
  and\ \bibinfo {author} {\bibfnamefont {S.}~\bibnamefont {Thomas}},\ }\href
  {\doibase https://doi.org/10.1016/0370-2693(94)90830-3} {\bibfield  {journal}
  {\bibinfo  {journal} {Physics Letters B}\ }\textbf {\bibinfo {volume}
  {320}},\ \bibinfo {pages} {99} (\bibinfo {year} {1994})}\BibitemShut
  {NoStop}%
\bibitem [{\citenamefont {Sigurdson}\ \emph {et~al.}(2004)\citenamefont
  {Sigurdson}, \citenamefont {Doran}, \citenamefont {Kurylov}, \citenamefont
  {Caldwell},\ and\ \citenamefont {Kamionkowski}}]{Sigurdson2004}%
  \BibitemOpen
  \bibfield  {author} {\bibinfo {author} {\bibfnamefont {K.}~\bibnamefont
  {Sigurdson}}, \bibinfo {author} {\bibfnamefont {M.}~\bibnamefont {Doran}},
  \bibinfo {author} {\bibfnamefont {A.}~\bibnamefont {Kurylov}}, \bibinfo
  {author} {\bibfnamefont {R.}~\bibnamefont {Caldwell}}, \ and\ \bibinfo
  {author} {\bibfnamefont {M.}~\bibnamefont {Kamionkowski}},\ }\href {\doibase
  10.1103/PhysRevD.70.083501} {\bibfield  {journal} {\bibinfo  {journal} {Phys.
  Rev. D}\ }\textbf {\bibinfo {volume} {70}} (\bibinfo {year} {2004}),\
  10.1103/PhysRevD.70.083501}\BibitemShut {NoStop}%
\bibitem [{\citenamefont {Fortin}\ and\ \citenamefont
  {Tait}(2012)}]{osti_1104185}%
  \BibitemOpen
  \bibfield  {author} {\bibinfo {author} {\bibfnamefont {J.-F.}\ \bibnamefont
  {Fortin}}\ and\ \bibinfo {author} {\bibfnamefont {T.~M.~P.}\ \bibnamefont
  {Tait}},\ }\href {\doibase 10.1103/PhysRevD.85.063506} {\bibfield  {journal}
  {\bibinfo  {journal} {Physical Review D}\ }\textbf {\bibinfo {volume} {85}}
  (\bibinfo {year} {2012}),\ 10.1103/PhysRevD.85.063506}\BibitemShut {NoStop}%
\bibitem [{\citenamefont {Banks}\ \emph {et~al.}(2010)\citenamefont {Banks},
  \citenamefont {Fortin},\ and\ \citenamefont {Thomas}}]{Banks:2010eh}%
  \BibitemOpen
  \bibfield  {author} {\bibinfo {author} {\bibfnamefont {T.}~\bibnamefont
  {Banks}}, \bibinfo {author} {\bibfnamefont {J.-F.}\ \bibnamefont {Fortin}}, \
  and\ \bibinfo {author} {\bibfnamefont {S.}~\bibnamefont {Thomas}},\
  }\href@noop {} {\  (\bibinfo {year} {2010})},\ \Eprint
  {http://arxiv.org/abs/1007.5515} {arXiv:1007.5515 [hep-ph]} \BibitemShut
  {NoStop}%
\bibitem [{\citenamefont {Antipin}\ \emph {et~al.}(2015)\citenamefont
  {Antipin}, \citenamefont {Redi}, \citenamefont {Strumia},\ and\ \citenamefont
  {Vigiani}}]{Antipin2015}%
  \BibitemOpen
  \bibfield  {author} {\bibinfo {author} {\bibfnamefont {O.}~\bibnamefont
  {Antipin}}, \bibinfo {author} {\bibfnamefont {M.}~\bibnamefont {Redi}},
  \bibinfo {author} {\bibfnamefont {A.}~\bibnamefont {Strumia}}, \ and\
  \bibinfo {author} {\bibfnamefont {E.}~\bibnamefont {Vigiani}},\ }\href
  {\doibase 10.1007/JHEP07(2015)039} {\bibfield  {journal} {\bibinfo  {journal}
  {Journal of High Energy Physics}\ }\textbf {\bibinfo {volume} {2015}}
  (\bibinfo {year} {2015}),\ 10.1007/JHEP07(2015)039}\BibitemShut {NoStop}%
\bibitem [{\citenamefont {Gresham}\ and\ \citenamefont
  {Zurek}(2014{\natexlab{b}})}]{PhysRevD.89.016017}%
  \BibitemOpen
  \bibfield  {author} {\bibinfo {author} {\bibfnamefont {M.~I.}\ \bibnamefont
  {Gresham}}\ and\ \bibinfo {author} {\bibfnamefont {K.~M.}\ \bibnamefont
  {Zurek}},\ }\href {\doibase 10.1103/PhysRevD.89.016017} {\bibfield  {journal}
  {\bibinfo  {journal} {Phys. Rev. D}\ }\textbf {\bibinfo {volume} {89}},\
  \bibinfo {pages} {016017} (\bibinfo {year} {2014}{\natexlab{b}})}\BibitemShut
  {NoStop}%
\bibitem [{\citenamefont {Fitzpatrick}\ and\ \citenamefont
  {Zurek}(2010)}]{PhysRevD.82.075004}%
  \BibitemOpen
  \bibfield  {author} {\bibinfo {author} {\bibfnamefont {A.~L.}\ \bibnamefont
  {Fitzpatrick}}\ and\ \bibinfo {author} {\bibfnamefont {K.~M.}\ \bibnamefont
  {Zurek}},\ }\href {\doibase 10.1103/PhysRevD.82.075004} {\bibfield  {journal}
  {\bibinfo  {journal} {Phys. Rev. D}\ }\textbf {\bibinfo {volume} {82}},\
  \bibinfo {pages} {075004} (\bibinfo {year} {2010})}\BibitemShut {NoStop}%
\bibitem [{\citenamefont {Barger}\ \emph {et~al.}(2011)\citenamefont {Barger},
  \citenamefont {Keung},\ and\ \citenamefont {Marfatia}}]{BARGER201174}%
  \BibitemOpen
  \bibfield  {author} {\bibinfo {author} {\bibfnamefont {V.}~\bibnamefont
  {Barger}}, \bibinfo {author} {\bibfnamefont {W.-Y.}\ \bibnamefont {Keung}}, \
  and\ \bibinfo {author} {\bibfnamefont {D.}~\bibnamefont {Marfatia}},\ }\href
  {\doibase https://doi.org/10.1016/j.physletb.2010.12.008} {\bibfield
  {journal} {\bibinfo  {journal} {Physics Letters B}\ }\textbf {\bibinfo
  {volume} {696}},\ \bibinfo {pages} {74} (\bibinfo {year} {2011})}\BibitemShut
  {NoStop}%
\bibitem [{\citenamefont {Nobile}\ \emph {et~al.}(2014)\citenamefont {Nobile},
  \citenamefont {Gelmini}, \citenamefont {Gondolo},\ and\ \citenamefont
  {Huh}}]{Nobile_2014}%
  \BibitemOpen
  \bibfield  {author} {\bibinfo {author} {\bibfnamefont {E.~D.}\ \bibnamefont
  {Nobile}}, \bibinfo {author} {\bibfnamefont {G.~B.}\ \bibnamefont {Gelmini}},
  \bibinfo {author} {\bibfnamefont {P.}~\bibnamefont {Gondolo}}, \ and\
  \bibinfo {author} {\bibfnamefont {J.-H.}\ \bibnamefont {Huh}},\ }\href
  {\doibase 10.1088/1475-7516/2014/06/002} {\bibfield  {journal} {\bibinfo
  {journal} {Journal of Cosmology and Astroparticle Physics}\ }\textbf
  {\bibinfo {volume} {2014}},\ \bibinfo {pages} {002} (\bibinfo {year}
  {2014})}\BibitemShut {NoStop}%
\bibitem [{\citenamefont {Gamboa}\ \emph {et~al.}(2020)\citenamefont {Gamboa},
  \citenamefont {M\'endez},\ and\ \citenamefont {Tapia}}]{PhysRevD.101.015013}%
  \BibitemOpen
  \bibfield  {author} {\bibinfo {author} {\bibfnamefont {J.}~\bibnamefont
  {Gamboa}}, \bibinfo {author} {\bibfnamefont {F.}~\bibnamefont {M\'endez}}, \
  and\ \bibinfo {author} {\bibfnamefont {N.}~\bibnamefont {Tapia}},\ }\href
  {\doibase 10.1103/PhysRevD.101.015013} {\bibfield  {journal} {\bibinfo
  {journal} {Phys. Rev. D}\ }\textbf {\bibinfo {volume} {101}},\ \bibinfo
  {pages} {015013} (\bibinfo {year} {2020})}\BibitemShut {NoStop}%
\bibitem [{\citenamefont {Arias}\ \emph {et~al.}(2019)\citenamefont {Arias},
  \citenamefont {Gamboa},\ and\ \citenamefont {Tapia}}]{ARIAS201917}%
  \BibitemOpen
  \bibfield  {author} {\bibinfo {author} {\bibfnamefont {P.}~\bibnamefont
  {Arias}}, \bibinfo {author} {\bibfnamefont {J.}~\bibnamefont {Gamboa}}, \
  and\ \bibinfo {author} {\bibfnamefont {N.}~\bibnamefont {Tapia}},\ }\href
  {\doibase https://doi.org/10.1016/j.physletb.2019.02.021} {\bibfield
  {journal} {\bibinfo  {journal} {Physics Letters B}\ }\textbf {\bibinfo
  {volume} {791}},\ \bibinfo {pages} {17} (\bibinfo {year} {2019})}\BibitemShut
  {NoStop}%
\bibitem [{\citenamefont {Kang}\ \emph {et~al.}(2018)\citenamefont {Kang},
  \citenamefont {Scopel}, \citenamefont {Tomar}, \citenamefont {Yoon},\ and\
  \citenamefont {Gondolo}}]{Kang_2018}%
  \BibitemOpen
  \bibfield  {author} {\bibinfo {author} {\bibfnamefont {S.}~\bibnamefont
  {Kang}}, \bibinfo {author} {\bibfnamefont {S.}~\bibnamefont {Scopel}},
  \bibinfo {author} {\bibfnamefont {G.}~\bibnamefont {Tomar}}, \bibinfo
  {author} {\bibfnamefont {J.-H.}\ \bibnamefont {Yoon}}, \ and\ \bibinfo
  {author} {\bibfnamefont {P.}~\bibnamefont {Gondolo}},\ }\href {\doibase
  10.1088/1475-7516/2018/11/040} {\bibfield  {journal} {\bibinfo  {journal}
  {Journal of Cosmology and Astroparticle Physics}\ }\textbf {\bibinfo {volume}
  {2018}},\ \bibinfo {pages} {040} (\bibinfo {year} {2018})}\BibitemShut
  {NoStop}%
\bibitem [{\citenamefont {Fl\'orez}\ \emph {et~al.}(2019)\citenamefont
  {Fl\'orez}, \citenamefont {Gurrola}, \citenamefont {Johns}, \citenamefont
  {Maruri}, \citenamefont {Sheldon}, \citenamefont {Sinha},\ and\ \citenamefont
  {Starko}}]{PhysRevD.100.016017}%
  \BibitemOpen
  \bibfield  {author} {\bibinfo {author} {\bibfnamefont {A.}~\bibnamefont
  {Fl\'orez}}, \bibinfo {author} {\bibfnamefont {A.}~\bibnamefont {Gurrola}},
  \bibinfo {author} {\bibfnamefont {W.}~\bibnamefont {Johns}}, \bibinfo
  {author} {\bibfnamefont {J.}~\bibnamefont {Maruri}}, \bibinfo {author}
  {\bibfnamefont {P.}~\bibnamefont {Sheldon}}, \bibinfo {author} {\bibfnamefont
  {K.}~\bibnamefont {Sinha}}, \ and\ \bibinfo {author} {\bibfnamefont {S.~R.}\
  \bibnamefont {Starko}},\ }\href {\doibase 10.1103/PhysRevD.100.016017}
  {\bibfield  {journal} {\bibinfo  {journal} {Phys. Rev. D}\ }\textbf {\bibinfo
  {volume} {100}},\ \bibinfo {pages} {016017} (\bibinfo {year}
  {2019})}\BibitemShut {NoStop}%
\bibitem [{\citenamefont {Vento}(2021)}]{Vento2021}%
  \BibitemOpen
  \bibfield  {author} {\bibinfo {author} {\bibfnamefont {V.}~\bibnamefont
  {Vento}},\ }\href {\doibase 10.1140/epjc/s10052-021-09027-6} {\bibfield
  {journal} {\bibinfo  {journal} {The European Physical Journal C}\ }\textbf
  {\bibinfo {volume} {81}} (\bibinfo {year} {2021}),\
  10.1140/epjc/s10052-021-09027-6}\BibitemShut {NoStop}%
\bibitem [{\citenamefont {Hisano}\ \emph {et~al.}(2020)\citenamefont {Hisano},
  \citenamefont {Ibarra},\ and\ \citenamefont {Nagai}}]{Hisano_2020}%
  \BibitemOpen
  \bibfield  {author} {\bibinfo {author} {\bibfnamefont {J.}~\bibnamefont
  {Hisano}}, \bibinfo {author} {\bibfnamefont {A.}~\bibnamefont {Ibarra}}, \
  and\ \bibinfo {author} {\bibfnamefont {R.}~\bibnamefont {Nagai}},\ }\href
  {\doibase 10.1088/1475-7516/2020/10/015} {\bibfield  {journal} {\bibinfo
  {journal} {Journal of Cosmology and Astroparticle Physics}\ }\textbf
  {\bibinfo {volume} {2020}},\ \bibinfo {pages} {015} (\bibinfo {year}
  {2020})}\BibitemShut {NoStop}%
\bibitem [{\citenamefont {Hambye}\ and\ \citenamefont {Xu}(2021)}]{Hambye2021}%
  \BibitemOpen
  \bibfield  {author} {\bibinfo {author} {\bibfnamefont {T.}~\bibnamefont
  {Hambye}}\ and\ \bibinfo {author} {\bibfnamefont {X.-J.}\ \bibnamefont
  {Xu}},\ }\href {https://doi.org/10.1007/JHEP11(2021)156} {\bibfield
  {journal} {\bibinfo  {journal} {Journal of High Energy Physics}\ } (\bibinfo
  {year} {2021})}\BibitemShut {NoStop}%
\bibitem [{\citenamefont {Bogorad}\ and\ \citenamefont
  {Toro}(2021)}]{Bogorad2021}%
  \BibitemOpen
  \bibfield  {author} {\bibinfo {author} {\bibfnamefont {Z.}~\bibnamefont
  {Bogorad}}\ and\ \bibinfo {author} {\bibfnamefont {N.}~\bibnamefont {Toro}},\
  }\href@noop {} {\  (\bibinfo {year} {2021})},\ \Eprint
  {http://arxiv.org/abs/2112.11476} {arXiv:2112.11476 [hep-ph]} \BibitemShut
  {NoStop}%
\bibitem [{\citenamefont {Li}\ and\ \citenamefont {Liu}(2022)}]{Li_2022}%
  \BibitemOpen
  \bibfield  {author} {\bibinfo {author} {\bibfnamefont {Q.}~\bibnamefont
  {Li}}\ and\ \bibinfo {author} {\bibfnamefont {Z.}~\bibnamefont {Liu}},\
  }\href {\doibase 10.1088/1674-1137/ac3d2b} {\bibfield  {journal} {\bibinfo
  {journal} {Chinese Physics C}\ }\textbf {\bibinfo {volume} {46}},\ \bibinfo
  {pages} {045102} (\bibinfo {year} {2022})}\BibitemShut {NoStop}%
\bibitem [{\citenamefont {Aboubrahim}\ \emph {et~al.}(2021)\citenamefont
  {Aboubrahim}, \citenamefont {Nath},\ and\ \citenamefont
  {Wang}}]{Aboubrahim2021}%
  \BibitemOpen
  \bibfield  {author} {\bibinfo {author} {\bibfnamefont {A.}~\bibnamefont
  {Aboubrahim}}, \bibinfo {author} {\bibfnamefont {P.}~\bibnamefont {Nath}}, \
  and\ \bibinfo {author} {\bibfnamefont {Z.-Y.}\ \bibnamefont {Wang}},\ }\href
  {\doibase 10.1007/JHEP12(2021)148} {\bibfield  {journal} {\bibinfo  {journal}
  {Journal of High Energy Physics}\ }\textbf {\bibinfo {volume} {2021}},\
  \bibinfo {pages} {148} (\bibinfo {year} {2021})}\BibitemShut {NoStop}%
\bibitem [{\citenamefont {Budker}\ \emph
  {et~al.}(2022{\natexlab{a}})\citenamefont {Budker}, \citenamefont {Graham},
  \citenamefont {Ramani}, \citenamefont {Schmidt-Kaler}, \citenamefont
  {Smorra},\ and\ \citenamefont {Ulmer}}]{osti_1847202}%
  \BibitemOpen
  \bibfield  {author} {\bibinfo {author} {\bibfnamefont {D.}~\bibnamefont
  {Budker}}, \bibinfo {author} {\bibfnamefont {P.~W.}\ \bibnamefont {Graham}},
  \bibinfo {author} {\bibfnamefont {H.}~\bibnamefont {Ramani}}, \bibinfo
  {author} {\bibfnamefont {F.}~\bibnamefont {Schmidt-Kaler}}, \bibinfo {author}
  {\bibfnamefont {C.}~\bibnamefont {Smorra}}, \ and\ \bibinfo {author}
  {\bibfnamefont {S.}~\bibnamefont {Ulmer}},\ }\href {\doibase
  10.1103/PRXQuantum.3.010330} {\bibfield  {journal} {\bibinfo  {journal} {PRX
  Quantum}\ }\textbf {\bibinfo {volume} {3}} (\bibinfo {year}
  {2022}{\natexlab{a}}),\ 10.1103/PRXQuantum.3.010330}\BibitemShut {NoStop}%
\bibitem [{\citenamefont {Jaeckel}\ and\ \citenamefont
  {Schenk}(2021)}]{PhysRevD.103.103523}%
  \BibitemOpen
  \bibfield  {author} {\bibinfo {author} {\bibfnamefont {J.}~\bibnamefont
  {Jaeckel}}\ and\ \bibinfo {author} {\bibfnamefont {S.}~\bibnamefont
  {Schenk}},\ }\href {\doibase 10.1103/PhysRevD.103.103523} {\bibfield
  {journal} {\bibinfo  {journal} {Phys. Rev. D}\ }\textbf {\bibinfo {volume}
  {103}},\ \bibinfo {pages} {103523} (\bibinfo {year} {2021})}\BibitemShut
  {NoStop}%
\bibitem [{\citenamefont {Mu{\~{n}}oz}\ and\ \citenamefont
  {Loeb}(2018)}]{Munnoz2018}%
  \BibitemOpen
  \bibfield  {author} {\bibinfo {author} {\bibfnamefont {J.~B.}\ \bibnamefont
  {Mu{\~{n}}oz}}\ and\ \bibinfo {author} {\bibfnamefont {A.}~\bibnamefont
  {Loeb}},\ }\href {\doibase 10.1038/s41586-018-0151-x} {\bibfield  {journal}
  {\bibinfo  {journal} {Nature}\ }\textbf {\bibinfo {volume} {557}},\ \bibinfo
  {pages} {684} (\bibinfo {year} {2018})}\BibitemShut {NoStop}%
\bibitem [{\citenamefont {Arina}\ \emph {et~al.}(2021)\citenamefont {Arina},
  \citenamefont {Cheek}, \citenamefont {Mimasu},\ and\ \citenamefont
  {Pagani}}]{Arina2021}%
  \BibitemOpen
  \bibfield  {author} {\bibinfo {author} {\bibfnamefont {C.}~\bibnamefont
  {Arina}}, \bibinfo {author} {\bibfnamefont {A.}~\bibnamefont {Cheek}},
  \bibinfo {author} {\bibfnamefont {K.}~\bibnamefont {Mimasu}}, \ and\ \bibinfo
  {author} {\bibfnamefont {L.}~\bibnamefont {Pagani}},\ }\href {\doibase
  10.1140/epjc/s10052-021-09010-1} {\bibfield  {journal} {\bibinfo  {journal}
  {The European Physical Journal C}\ }\textbf {\bibinfo {volume} {81}},\
  \bibinfo {pages} {223} (\bibinfo {year} {2021})}\BibitemShut {NoStop}%
\bibitem [{\citenamefont {Smith}(2012)}]{Smith:2012fq}%
  \BibitemOpen
  \bibfield  {author} {\bibinfo {author} {\bibfnamefont {N.~J.~T.}\
  \bibnamefont {Smith}},\ }\href {\doibase 10.1140/epjp/i2012-12108-9}
  {\bibfield  {journal} {\bibinfo  {journal} {Eur. Phys. J. Plus}\ }\textbf
  {\bibinfo {volume} {127}},\ \bibinfo {pages} {108} (\bibinfo {year}
  {2012})}\BibitemShut {NoStop}%
\bibitem [{\citenamefont {Cowan}\ \emph {et~al.}(2011)\citenamefont {Cowan},
  \citenamefont {Cranmer}, \citenamefont {Gross},\ and\ \citenamefont
  {Vitells}}]{PLR_2011}%
  \BibitemOpen
  \bibfield  {author} {\bibinfo {author} {\bibfnamefont {G.}~\bibnamefont
  {Cowan}}, \bibinfo {author} {\bibfnamefont {K.}~\bibnamefont {Cranmer}},
  \bibinfo {author} {\bibfnamefont {E.}~\bibnamefont {Gross}}, \ and\ \bibinfo
  {author} {\bibfnamefont {O.}~\bibnamefont {Vitells}},\ }\href {\doibase
  10.1140/epjc/s10052-011-1554-0} {\bibfield  {journal} {\bibinfo  {journal}
  {The European Physical Journal C}\ }\textbf {\bibinfo {volume} {71}}
  (\bibinfo {year} {2011}),\ 10.1140/epjc/s10052-011-1554-0}\BibitemShut
  {NoStop}%
\bibitem [{\citenamefont {Foreman-Mackey}\ \emph {et~al.}(2013)\citenamefont
  {Foreman-Mackey}, \citenamefont {Hogg}, \citenamefont {Lang},\ and\
  \citenamefont {Goodman}}]{Foreman_Mackey_2013}%
  \BibitemOpen
  \bibfield  {author} {\bibinfo {author} {\bibfnamefont {D.}~\bibnamefont
  {Foreman-Mackey}}, \bibinfo {author} {\bibfnamefont {D.~W.}\ \bibnamefont
  {Hogg}}, \bibinfo {author} {\bibfnamefont {D.}~\bibnamefont {Lang}}, \ and\
  \bibinfo {author} {\bibfnamefont {J.}~\bibnamefont {Goodman}},\ }\href
  {\doibase 10.1086/670067} {\bibfield  {journal} {\bibinfo  {journal}
  {Publications of the Astronomical Society of the Pacific}\ }\textbf {\bibinfo
  {volume} {125}},\ \bibinfo {pages} {306} (\bibinfo {year}
  {2013})}\BibitemShut {NoStop}%
\bibitem [{\citenamefont {Durnford}\ and\ \citenamefont
  {Piro}(2022)}]{Durnford_2022}%
  \BibitemOpen
  \bibfield  {author} {\bibinfo {author} {\bibfnamefont {D.}~\bibnamefont
  {Durnford}}\ and\ \bibinfo {author} {\bibfnamefont {M.-C.}\ \bibnamefont
  {Piro}},\ }\href {\doibase 10.1088/1748-0221/17/01/c01030} {\bibfield
  {journal} {\bibinfo  {journal} {Journal of Instrumentation}\ }\textbf
  {\bibinfo {volume} {17}},\ \bibinfo {pages} {C01030} (\bibinfo {year}
  {2022})}\BibitemShut {NoStop}%
\bibitem [{\citenamefont {Lewin}\ and\ \citenamefont
  {Smith}(1996)}]{LEWIN199687}%
  \BibitemOpen
  \bibfield  {author} {\bibinfo {author} {\bibfnamefont {J.}~\bibnamefont
  {Lewin}}\ and\ \bibinfo {author} {\bibfnamefont {P.}~\bibnamefont {Smith}},\
  }\href {\doibase https://doi.org/10.1016/S0927-6505(96)00047-3} {\bibfield
  {journal} {\bibinfo  {journal} {Astroparticle Physics}\ }\textbf {\bibinfo
  {volume} {6}},\ \bibinfo {pages} {87} (\bibinfo {year} {1996})}\BibitemShut
  {NoStop}%
\bibitem [{\citenamefont {Fan}\ \emph {et~al.}(2010)\citenamefont {Fan},
  \citenamefont {Reece},\ and\ \citenamefont {Wang}}]{Fan_2010}%
  \BibitemOpen
  \bibfield  {author} {\bibinfo {author} {\bibfnamefont {J.}~\bibnamefont
  {Fan}}, \bibinfo {author} {\bibfnamefont {M.}~\bibnamefont {Reece}}, \ and\
  \bibinfo {author} {\bibfnamefont {L.-T.}\ \bibnamefont {Wang}},\ }\href
  {\doibase 10.1088/1475-7516/2010/11/042} {\bibfield  {journal} {\bibinfo
  {journal} {Journal of Cosmology and Astroparticle Physics}\ }\textbf
  {\bibinfo {volume} {2010}},\ \bibinfo {pages} {042} (\bibinfo {year}
  {2010})}\BibitemShut {NoStop}%
\bibitem [{\citenamefont {Fitzpatrick}\ \emph {et~al.}(2013)\citenamefont
  {Fitzpatrick}, \citenamefont {Haxton}, \citenamefont {Katz}, \citenamefont
  {Lubbers},\ and\ \citenamefont {Xu}}]{fitzpatrick_effective_2013}%
  \BibitemOpen
  \bibfield  {author} {\bibinfo {author} {\bibfnamefont {A.~L.}\ \bibnamefont
  {Fitzpatrick}}, \bibinfo {author} {\bibfnamefont {W.}~\bibnamefont {Haxton}},
  \bibinfo {author} {\bibfnamefont {E.}~\bibnamefont {Katz}}, \bibinfo {author}
  {\bibfnamefont {N.}~\bibnamefont {Lubbers}}, \ and\ \bibinfo {author}
  {\bibfnamefont {Y.}~\bibnamefont {Xu}},\ }\href {\doibase
  10.1088/1475-7516/2013/02/004} {\bibfield  {journal} {\bibinfo  {journal} {J.
  Cosmol. Astropart. Phys.}\ }\textbf {\bibinfo {volume} {2013}},\ \bibinfo
  {pages} {004} (\bibinfo {year} {2013})}\BibitemShut {NoStop}%
\bibitem [{\citenamefont {Anand}\ \emph {et~al.}(2014)\citenamefont {Anand},
  \citenamefont {Fitzpatrick},\ and\ \citenamefont
  {Haxton}}]{PhysRevC.89.065501}%
  \BibitemOpen
  \bibfield  {author} {\bibinfo {author} {\bibfnamefont {N.}~\bibnamefont
  {Anand}}, \bibinfo {author} {\bibfnamefont {A.~L.}\ \bibnamefont
  {Fitzpatrick}}, \ and\ \bibinfo {author} {\bibfnamefont {W.~C.}\ \bibnamefont
  {Haxton}},\ }\href {\doibase 10.1103/PhysRevC.89.065501} {\bibfield
  {journal} {\bibinfo  {journal} {Phys. Rev. C}\ }\textbf {\bibinfo {volume}
  {89}},\ \bibinfo {pages} {065501} (\bibinfo {year} {2014})}\BibitemShut
  {NoStop}%
\bibitem [{\citenamefont {Dent}\ \emph {et~al.}(2015)\citenamefont {Dent},
  \citenamefont {Krauss}, \citenamefont {Newstead},\ and\ \citenamefont
  {Sabharwal}}]{PhysRevD.92.063515}%
  \BibitemOpen
  \bibfield  {author} {\bibinfo {author} {\bibfnamefont {J.~B.}\ \bibnamefont
  {Dent}}, \bibinfo {author} {\bibfnamefont {L.~M.}\ \bibnamefont {Krauss}},
  \bibinfo {author} {\bibfnamefont {J.~L.}\ \bibnamefont {Newstead}}, \ and\
  \bibinfo {author} {\bibfnamefont {S.}~\bibnamefont {Sabharwal}},\ }\href
  {\doibase 10.1103/PhysRevD.92.063515} {\bibfield  {journal} {\bibinfo
  {journal} {Phys. Rev. D}\ }\textbf {\bibinfo {volume} {92}},\ \bibinfo
  {pages} {063515} (\bibinfo {year} {2015})}\BibitemShut {NoStop}%
\bibitem [{\citenamefont {Kavanagh}\ and\ \citenamefont
  {Edwards}(2018)}]{WIMpy-code}%
  \BibitemOpen
  \bibfield  {author} {\bibinfo {author} {\bibfnamefont {B.~J.}\ \bibnamefont
  {Kavanagh}}\ and\ \bibinfo {author} {\bibfnamefont {T.~D.~P.}\ \bibnamefont
  {Edwards}},\ }\href@noop {} {\enquote {\bibinfo {title}
  {\textnormal{WIMpy\_NREFT v1.1 [Computer Software]},
  \href{https://doi.org/10.5281/zenodo.1230503}{\textnormal{doi:10.5281/zenodo.1230503}}\textnormal{.
  Available at }\url{https://github.com/bradkav/WIMpy_NREFT}},}\ } (\bibinfo
  {year} {2018})\BibitemShut {NoStop}%
\bibitem [{\citenamefont {Gluscevic}\ \emph {et~al.}(2015)\citenamefont
  {Gluscevic}, \citenamefont {Gresham}, \citenamefont {McDermott},
  \citenamefont {Peter},\ and\ \citenamefont {Zurek}}]{Gluscevic_2015}%
  \BibitemOpen
  \bibfield  {author} {\bibinfo {author} {\bibfnamefont {V.}~\bibnamefont
  {Gluscevic}}, \bibinfo {author} {\bibfnamefont {M.~I.}\ \bibnamefont
  {Gresham}}, \bibinfo {author} {\bibfnamefont {S.~D.}\ \bibnamefont
  {McDermott}}, \bibinfo {author} {\bibfnamefont {A.~H.}\ \bibnamefont
  {Peter}}, \ and\ \bibinfo {author} {\bibfnamefont {K.~M.}\ \bibnamefont
  {Zurek}},\ }\href {\doibase 10.1088/1475-7516/2015/12/057} {\bibfield
  {journal} {\bibinfo  {journal} {Journal of Cosmology and Astroparticle
  Physics}\ }\textbf {\bibinfo {volume} {2015}},\ \bibinfo {pages} {057}
  (\bibinfo {year} {2015})}\BibitemShut {NoStop}%
\bibitem [{\citenamefont {Gluscevic}\ and\ \citenamefont
  {McDermott}(2015)}]{dmdd}%
  \BibitemOpen
  \bibfield  {author} {\bibinfo {author} {\bibfnamefont {V.}~\bibnamefont
  {Gluscevic}}\ and\ \bibinfo {author} {\bibfnamefont {S.~D.}\ \bibnamefont
  {McDermott}},\ }\href {https://ui.adsabs.harvard.edu/abs/2015ascl.soft06002G}
  {\bibfield  {journal} {\bibinfo  {journal} {Astrophysics Source Code
  Library}\ ,\ \bibinfo {pages} {ascl:1506.002}} (\bibinfo {year}
  {2015})}\BibitemShut {NoStop}%
\bibitem [{\citenamefont {Ho}\ and\ \citenamefont
  {Scherrer}(2013)}]{HO2013341}%
  \BibitemOpen
  \bibfield  {author} {\bibinfo {author} {\bibfnamefont {C.~M.}\ \bibnamefont
  {Ho}}\ and\ \bibinfo {author} {\bibfnamefont {R.~J.}\ \bibnamefont
  {Scherrer}},\ }\href {\doibase
  https://doi.org/10.1016/j.physletb.2013.04.039} {\bibfield  {journal}
  {\bibinfo  {journal} {Physics Letters B}\ }\textbf {\bibinfo {volume}
  {722}},\ \bibinfo {pages} {341} (\bibinfo {year} {2013})}\BibitemShut
  {NoStop}%
\bibitem [{\citenamefont {Buch}\ \emph {et~al.}(2020)\citenamefont {Buch},
  \citenamefont {Fan},\ and\ \citenamefont {Leung}}]{PhysRevD.101.063026}%
  \BibitemOpen
  \bibfield  {author} {\bibinfo {author} {\bibfnamefont {J.}~\bibnamefont
  {Buch}}, \bibinfo {author} {\bibfnamefont {J.}~\bibnamefont {Fan}}, \ and\
  \bibinfo {author} {\bibfnamefont {J.~S.~C.}\ \bibnamefont {Leung}},\ }\href
  {\doibase 10.1103/PhysRevD.101.063026} {\bibfield  {journal} {\bibinfo
  {journal} {Phys. Rev. D}\ }\textbf {\bibinfo {volume} {101}},\ \bibinfo
  {pages} {063026} (\bibinfo {year} {2020})}\BibitemShut {NoStop}%
\bibitem [{\citenamefont {Adhikari}\ \emph {et~al.}(2020)\citenamefont
  {Adhikari} \emph {et~al.}}]{PhysRevD.102.082001}%
  \BibitemOpen
  \bibfield  {author} {\bibinfo {author} {\bibfnamefont {P.}~\bibnamefont
  {Adhikari}} \emph {et~al.} (\bibinfo {collaboration} {DEAP Collaboration}),\
  }\href {\doibase 10.1103/PhysRevD.102.082001} {\bibfield  {journal} {\bibinfo
   {journal} {Phys. Rev. D}\ }\textbf {\bibinfo {volume} {102}},\ \bibinfo
  {pages} {082001} (\bibinfo {year} {2020})}\BibitemShut {NoStop}%
\bibitem [{\citenamefont {Holdom}(1986)}]{HOLDOM1986196}%
  \BibitemOpen
  \bibfield  {author} {\bibinfo {author} {\bibfnamefont {B.}~\bibnamefont
  {Holdom}},\ }\href {\doibase https://doi.org/10.1016/0370-2693(86)91377-8}
  {\bibfield  {journal} {\bibinfo  {journal} {Physics Letters B}\ }\textbf
  {\bibinfo {volume} {166}},\ \bibinfo {pages} {196} (\bibinfo {year}
  {1986})}\BibitemShut {NoStop}%
\bibitem [{\citenamefont {Izaguirre}\ and\ \citenamefont
  {Yavin}(2015)}]{PhysRevD.92.035014}%
  \BibitemOpen
  \bibfield  {author} {\bibinfo {author} {\bibfnamefont {E.}~\bibnamefont
  {Izaguirre}}\ and\ \bibinfo {author} {\bibfnamefont {I.}~\bibnamefont
  {Yavin}},\ }\href {\doibase 10.1103/PhysRevD.92.035014} {\bibfield  {journal}
  {\bibinfo  {journal} {Phys. Rev. D}\ }\textbf {\bibinfo {volume} {92}},\
  \bibinfo {pages} {035014} (\bibinfo {year} {2015})}\BibitemShut {NoStop}%
\bibitem [{\citenamefont {Feldman}\ \emph {et~al.}(2007)\citenamefont
  {Feldman}, \citenamefont {Liu},\ and\ \citenamefont
  {Nath}}]{PhysRevD.75.115001}%
  \BibitemOpen
  \bibfield  {author} {\bibinfo {author} {\bibfnamefont {D.}~\bibnamefont
  {Feldman}}, \bibinfo {author} {\bibfnamefont {Z.}~\bibnamefont {Liu}}, \ and\
  \bibinfo {author} {\bibfnamefont {P.}~\bibnamefont {Nath}},\ }\href {\doibase
  10.1103/PhysRevD.75.115001} {\bibfield  {journal} {\bibinfo  {journal} {Phys.
  Rev. D}\ }\textbf {\bibinfo {volume} {75}},\ \bibinfo {pages} {115001}
  (\bibinfo {year} {2007})}\BibitemShut {NoStop}%
\bibitem [{\citenamefont {Moreno}\ \emph {et~al.}(2019)\citenamefont {Moreno}
  \emph {et~al.}}]{LDMX:2019}%
  \BibitemOpen
  \bibfield  {author} {\bibinfo {author} {\bibfnamefont {O.}~\bibnamefont
  {Moreno}} \emph {et~al.} (\bibinfo {collaboration} {LDMX Collaboration}),\
  }in\ \href {https://ui.adsabs.harvard.edu/abs/2019APS..APRT04003M} {\emph
  {\bibinfo {booktitle} {APS April Meeting Abstracts}}},\ \bibinfo {series and
  number} {APS Meeting Abstracts}\ (\bibinfo {year} {2019})\ p.\ \bibinfo
  {pages} {T04.003}\BibitemShut {NoStop}%
\bibitem [{\citenamefont {Cline}\ \emph {et~al.}(2012)\citenamefont {Cline},
  \citenamefont {Liu},\ and\ \citenamefont {Xue}}]{PhysRevD.85.101302}%
  \BibitemOpen
  \bibfield  {author} {\bibinfo {author} {\bibfnamefont {J.~M.}\ \bibnamefont
  {Cline}}, \bibinfo {author} {\bibfnamefont {Z.}~\bibnamefont {Liu}}, \ and\
  \bibinfo {author} {\bibfnamefont {W.}~\bibnamefont {Xue}},\ }\href {\doibase
  10.1103/PhysRevD.85.101302} {\bibfield  {journal} {\bibinfo  {journal} {Phys.
  Rev. D}\ }\textbf {\bibinfo {volume} {85}},\ \bibinfo {pages} {101302}
  (\bibinfo {year} {2012})}\BibitemShut {NoStop}%
\bibitem [{\citenamefont {Budker}\ \emph
  {et~al.}(2022{\natexlab{b}})\citenamefont {Budker}, \citenamefont {Graham},
  \citenamefont {Ramani}, \citenamefont {Schmidt-Kaler}, \citenamefont
  {Smorra},\ and\ \citenamefont {Ulmer}}]{PRXQuantum.3.010330}%
  \BibitemOpen
  \bibfield  {author} {\bibinfo {author} {\bibfnamefont {D.}~\bibnamefont
  {Budker}}, \bibinfo {author} {\bibfnamefont {P.~W.}\ \bibnamefont {Graham}},
  \bibinfo {author} {\bibfnamefont {H.}~\bibnamefont {Ramani}}, \bibinfo
  {author} {\bibfnamefont {F.}~\bibnamefont {Schmidt-Kaler}}, \bibinfo {author}
  {\bibfnamefont {C.}~\bibnamefont {Smorra}}, \ and\ \bibinfo {author}
  {\bibfnamefont {S.}~\bibnamefont {Ulmer}},\ }\href {\doibase
  10.1103/PRXQuantum.3.010330} {\bibfield  {journal} {\bibinfo  {journal} {PRX
  Quantum}\ }\textbf {\bibinfo {volume} {3}},\ \bibinfo {pages} {010330}
  (\bibinfo {year} {2022}{\natexlab{b}})}\BibitemShut {NoStop}%
\bibitem [{\citenamefont {Magill}\ \emph
  {et~al.}(2019{\natexlab{a}})\citenamefont {Magill}, \citenamefont {Plestid},
  \citenamefont {Pospelov},\ and\ \citenamefont {Tsai}}]{Magill}%
  \BibitemOpen
  \bibfield  {author} {\bibinfo {author} {\bibfnamefont {G.}~\bibnamefont
  {Magill}}, \bibinfo {author} {\bibfnamefont {R.}~\bibnamefont {Plestid}},
  \bibinfo {author} {\bibfnamefont {M.}~\bibnamefont {Pospelov}}, \ and\
  \bibinfo {author} {\bibfnamefont {Y.-D.}\ \bibnamefont {Tsai}},\ }\href
  {\doibase 10.1103/PhysRevLett.122.071801} {\bibfield  {journal} {\bibinfo
  {journal} {Physical Review Letters}\ }\textbf {\bibinfo {volume} {122}}
  (\bibinfo {year} {2019}{\natexlab{a}}),\
  10.1103/PhysRevLett.122.071801}\BibitemShut {NoStop}%
\bibitem [{\citenamefont {Kelly}\ and\ \citenamefont
  {Tsai}(2019)}]{PhysRevD.100.015043}%
  \BibitemOpen
  \bibfield  {author} {\bibinfo {author} {\bibfnamefont {K.~J.}\ \bibnamefont
  {Kelly}}\ and\ \bibinfo {author} {\bibfnamefont {Y.-D.}\ \bibnamefont
  {Tsai}},\ }\href {\doibase 10.1103/PhysRevD.100.015043} {\bibfield  {journal}
  {\bibinfo  {journal} {Phys. Rev. D}\ }\textbf {\bibinfo {volume} {100}},\
  \bibinfo {pages} {015043} (\bibinfo {year} {2019})}\BibitemShut {NoStop}%
\bibitem [{\citenamefont {Plestid}\ \emph {et~al.}(2020)\citenamefont
  {Plestid}, \citenamefont {Takhistov}, \citenamefont {Tsai}, \citenamefont
  {Bringmann}, \citenamefont {Kusenko},\ and\ \citenamefont
  {Pospelov}}]{osti_1745080}%
  \BibitemOpen
  \bibfield  {author} {\bibinfo {author} {\bibfnamefont {R.}~\bibnamefont
  {Plestid}}, \bibinfo {author} {\bibfnamefont {V.}~\bibnamefont {Takhistov}},
  \bibinfo {author} {\bibfnamefont {Y.-D.}\ \bibnamefont {Tsai}}, \bibinfo
  {author} {\bibfnamefont {T.}~\bibnamefont {Bringmann}}, \bibinfo {author}
  {\bibfnamefont {A.}~\bibnamefont {Kusenko}}, \ and\ \bibinfo {author}
  {\bibfnamefont {M.}~\bibnamefont {Pospelov}},\ }\href {\doibase
  10.1103/PhysRevD.102.115032} {\bibfield  {journal} {\bibinfo  {journal}
  {Physical Review D}\ }\textbf {\bibinfo {volume} {102}} (\bibinfo {year}
  {2020}),\ 10.1103/PhysRevD.102.115032}\BibitemShut {NoStop}%
\bibitem [{\citenamefont {Liu}\ \emph {et~al.}(2019)\citenamefont {Liu},
  \citenamefont {Outmezguine}, \citenamefont {Redigolo},\ and\ \citenamefont
  {Volansky}}]{Hongwan}%
  \BibitemOpen
  \bibfield  {author} {\bibinfo {author} {\bibfnamefont {H.}~\bibnamefont
  {Liu}}, \bibinfo {author} {\bibfnamefont {N.}~\bibnamefont {Outmezguine}},
  \bibinfo {author} {\bibfnamefont {D.}~\bibnamefont {Redigolo}}, \ and\
  \bibinfo {author} {\bibfnamefont {T.}~\bibnamefont {Volansky}},\ }\href
  {\doibase 10.1103/PhysRevD.100.123011} {\bibfield  {journal} {\bibinfo
  {journal} {Physical Review D}\ }\textbf {\bibinfo {volume} {100}} (\bibinfo
  {year} {2019}),\ 10.1103/PhysRevD.100.123011}\BibitemShut {NoStop}%
\bibitem [{\citenamefont {Haas}\ \emph {et~al.}(2015)\citenamefont {Haas},
  \citenamefont {Hill}, \citenamefont {Izaguirre},\ and\ \citenamefont
  {Yavin}}]{HAAS2015117}%
  \BibitemOpen
  \bibfield  {author} {\bibinfo {author} {\bibfnamefont {A.}~\bibnamefont
  {Haas}}, \bibinfo {author} {\bibfnamefont {C.~S.}\ \bibnamefont {Hill}},
  \bibinfo {author} {\bibfnamefont {E.}~\bibnamefont {Izaguirre}}, \ and\
  \bibinfo {author} {\bibfnamefont {I.}~\bibnamefont {Yavin}},\ }\href
  {\doibase https://doi.org/10.1016/j.physletb.2015.04.062} {\bibfield
  {journal} {\bibinfo  {journal} {Physics Letters B}\ }\textbf {\bibinfo
  {volume} {746}},\ \bibinfo {pages} {117} (\bibinfo {year}
  {2015})}\BibitemShut {NoStop}%
\bibitem [{\citenamefont {Magill}\ \emph
  {et~al.}(2019{\natexlab{b}})\citenamefont {Magill}, \citenamefont {Plestid},
  \citenamefont {Pospelov},\ and\ \citenamefont
  {Tsai}}]{PhysRevLett.122.071801}%
  \BibitemOpen
  \bibfield  {author} {\bibinfo {author} {\bibfnamefont {G.}~\bibnamefont
  {Magill}}, \bibinfo {author} {\bibfnamefont {R.}~\bibnamefont {Plestid}},
  \bibinfo {author} {\bibfnamefont {M.}~\bibnamefont {Pospelov}}, \ and\
  \bibinfo {author} {\bibfnamefont {Y.-D.}\ \bibnamefont {Tsai}},\ }\href
  {\doibase 10.1103/PhysRevLett.122.071801} {\bibfield  {journal} {\bibinfo
  {journal} {Phys. Rev. Lett.}\ }\textbf {\bibinfo {volume} {122}},\ \bibinfo
  {pages} {071801} (\bibinfo {year} {2019}{\natexlab{b}})}\BibitemShut
  {NoStop}%
\end{thebibliography}%

\end{document}